\documentclass[fleqn,usenatbib]{mnras}

\usepackage[T1]{fontenc}

\DeclareRobustCommand{\VAN}[3]{#2}
\let\VANthebibliography\thebibliography
\def\thebibliography{\DeclareRobustCommand{\VAN}[3]{##3}\VANthebibliography}

\DeclareSymbolFont{matha}{OML}{txmi}{m}{it}
\DeclareMathSymbol{\vv}{\mathord}{matha}{29}

\newcommand{\HI}{H{\small{I}}}

\usepackage{graphicx}	
\usepackage{amsmath}	
\usepackage{amssymb}	
\usepackage{comment}
\usepackage{braket}
\usepackage{booktabs}

\title[Optimizing $21$-cm line stacking of galaxies]{Optimizing spectral stacking for 21-cm observations of galaxies: accuracy assessment and symmetrized stacking}

\author[F. Sinigaglia et al.]{\hspace{-0.15cm}
Francesco Sinigaglia,$^{1,2,3,4}$\thanks{Email: francesco.sinigaglia@phd.unipd.it}
Ed Elson,$^{5}$
Giulia Rodighiero,$^{1,2}$ and
Mattia Vaccari $^{6,7,8}$
\vspace{0.5cm}
\\
$^{1}$Department of Physics and Astronomy, Università degli Studi di Padova, Vicolo dell’Osservatorio 3, I-35122, Padova, Italy\\
$^{2}$INAF - Osservatorio Astronomico di Padova, Vicolo dell’Osservatorio 5, I-35122, Padova, Italy\\
$^{3}$Instituto de Astrof\'isica de Canarias, Calle Via L\'actea s/n, E-38205, La  Laguna, Tenerife, Spain\\
$^{4}$Departamento  de  Astrof\'isica, Universidad de La Laguna,  E-38206, La Laguna, Tenerife, Spain\\
$^{5}$Department of Physics and Astronomy, University of the Western Cape, Robert Sobukwe Rd, 7535 Bellville, Cape Town, South Africa\\
$^{6}$Inter-university Institute for Data Intensive Astronomy, Department of Physics and Astronomy, University of the Western Cape,\\ 7535 Bellville, Cape Town, South Africa\\
$^{7}$Inter-university Institute for Data Intensive Astronomy, Department of Astronomy, University of Cape Town,\\ 7701 Rondebosch, Cape Town, South Africa\\
$^{8}$INAF - Istituto di Radioastronomia, via Gobetti 101, 40129 Bologna, Italy
}

\date{Accepted XXX. Received YYY; in original form ZZZ}

\pubyear{2022}

\begin{document}
\label{firstpage}
\pagerange{\pageref{firstpage}--\pageref{lastpage}}
\maketitle

\begin{abstract}

We present an assessment of the accuracy of common operations performed in $21$-cm spectral line stacking experiments. To this end, we generate mock interferometric data surveying the 21-cm emission at frequency $1310<\nu<1420$ MHz ($0.005<z<0.084$) and covering an area $\sim 6$ deg$^2$ of the sky, mimicking the observational characteristics of real MeerKAT observations. We find that the primary beam correction accounts for just few per cent ($\sim8\%$ at 0 primary beam power, $\sim 3\%$ at 0.6 primary beam power) deviations from the true $M_{\rm HI}$ signal, and that weighting schemes based on noise properties provide unbiased results. On the contrary, weighting schemes based on distance can account for significant systematic mass differences when applied to a flux-limited sample ($\Delta M_{\rm HI}\sim 40-50\%$ in the studied case). We find no significant difference in the final $\braket{M_{\rm HI}}$ obtained when spectroscopic redshift uncertainties are accounted for in the stacking procedure ($ \Delta z\sim 0.00035$, i.e. $\Delta v \sim 100\,{\rm km\, s}^{-1}$). We also present a novel technique to increase the effective size of the galaxy sample by exploiting the geometric symmetries of galaxy cubelets, potentially enhancing the SNR by a factor $\sim\sqrt{2}$ when analyzing the final stacked spectrum (a factor 4 in a cubelet). This procedure is found to be robustly unbiased, while efficiently increasing the SNR, as expected. We argue that an appropriate framework employing detailed and realistic simulations is required to exploit upcoming datasets from SKA pathfinders in an accurate and reliable manner. 

\end{abstract}

\vspace{1cm}



\section{Introduction}

Understanding the distribution, cosmic evolution and phenomenology of neutral atomic hydrogen (\HI{} hereafter) is currently the subject of an intense theoretical and observational effort. In fact, \HI{} constitutes the fundamental component of H$_2$, representing therefore the raw fuel of star formation \citep[e.g. ][]{Blitz2006, Bigiel2008,Leroy2008,Krumholz2009,Glover2012,Sternberg2014,Diemer2019}. In this scenario, setting observational constraints on the \HI{} density evolution and content in galaxies is a task of paramount importance to develop a holistic picture of galaxy formation and evolution and understand how the availability of fresh gas through cosmological accretion can sustain star formation \citep{Keres2005,VanDeVoort2012,Conselice2013,SanchezAlmeida2014} and how processes such as feedback by active galactic nuclei (AGN) and supernovae, ram-pressure stripping and mergers can conversely lead to star formation quenching \citep[see e.g., ][and references therein]{Gabor2010,Pontzen2017,Bluck2020,Kalinova2021}.



To shed light onto these profound questions, pioneering works \citep[e.g.,][]{Haynes1984,Roberts1994} and several large-scale observational campaigns have mapped the distribution of \HI{} through the 21-cm hyperfine transition emission line at $z\sim 0$, such as the \HI{} Parkes All-Sky Survey \citep[HIPASS, ][]{Barnes2001,Meyer2004}, the Arecibo Fast Legacy ALFA Survey \citep[ALFALFA, ][]{Giovanelli2005} and the GALEX Arecibo SDSS Survey \citep[GASS, ][]{Catinella2010}. The emerging picture reveals that the \HI{} content of star-forming galaxies turns out to be tightly related to their stellar mass $M_*$ \citep[][]{Huang2012,Maddox2015,Parkash2018,Calette2018}, star formation rate \citep[][]{Huang2012,Feldmann2020}, optical colors \citep[][]{Huang2012}, disc size \citep[e.g. ][and references therein]{Wang2016} and magnitudes \citep[e.g. ][]{Denes2014}, among others. 


Direct \HI{} detections beyond the Local Universe ($z>0$) are however rare and challenging to perform, due to the intrinsic faintness of the 21-cm line in relation to the sensitivity of existing radio telescopes. Few blind deep observational efforts have been undertaken, as e.g. the Blind Ultra-Deep \HI{} Environmental Survey \citep[BUDHIES, ][]{Verheijen2007,Gogate2020}, the COSMOS \HI{} Large Extragalactic Survey \citep[CHILES, ][]{Hess2019} and the Arecibo Ultra-Deep Survey \citep[AUDS, ][]{Hoppman2015}, although they typically require extremely long integration times and are therefore not suited to investigate very large areas of the sky. The ongoing and planned \HI{} surveys conducted with the MeerKAT array -- i.e. the MeerKAT International GigaHertz Tiered Extragalactic Exploration \citep[MIGHTEE,][]{Jarvis2016,Maddox2021}, Looking At the Distant Universe with the MeerKAT Array survey \citep[LADUMA,][]{Blyth2016,Baker2018}, the MeerKAT Observations of Nearby Galactic Objects Observing Southern Emitters survey \citep[MHONGOOSE,][]{DeBlok2016}, and the MeerKAT Fornax Survey \citep[][]{Serra2016} -- and with the Australian SKA Pathifinder (ASKAP) -- i.e. the Widefield ASKAP L-band Legacy All-sky Blind surveY \citep[WALLABY,][]{Koribalski2020}, and the Deep Investigation of Neutral Gas Origins survey \citep[DINGO,][]{Meyer2009} --  promise to take a significant step forward in our understanding of the role of \HI{} in galaxy evolution out to $z\gtrsim 1$. Furthermore, surveys targeting intervening \HI{} absorption -- e.g. the The First Large Absorption Survey in \HI{} \citep[FLASH,][]{Allison2021} and the MeerKAT Absorption Line Survey \citep[MALS,][]{Gupta2016} -- offer a valuable alternative to stacking (discussed in what follows) to mine the \HI{} content in the Universe up to very high redshifts, where direct detections are not feasible.

Cosmological hydrodynamic simulations of galaxy formation and evolution manage to reproduce the majority of the results from observations in the Local Universe \citep[see e.g. ][and references therein]{Dave2020}, while the agreement with data at higher redshift has not been systematically assessed yet given the large uncertainties on the existing observational constraints. 

To overcome the 21-cm line faintness problem, a common technique employed in literature is the so-called \textit{spectral line stacking}. This consists in co-adding spectra centred on the rest-frame frequency of the target spectral line (individually undetected in each spectrum) to obtain a final signal above the noise level and perform an average spectral line detection, at the expense of the information on the underlying line flux emitted by individual galaxies. This method has recently grown in popularity and has been adopted to probe the presence and abundance of \HI{} in galaxy clusters \citep{Zwaan2000,Chengalur2001,Lah2009,Healy2021} to investigate scaling relations \citep{Fabello2011b,Fabello2012a,Brown2015,Gereb2015,Brown2017}, the \HI{} mass function \citep{Pan2020}, the $M_{\rm HI}$ content of AGN host galaxies \citep{Fabello2011b,Gereb2013,Gereb2015}, the baryonic Tully-Fisher relation \citep{Meyer2016}, the \HI{} cosmic density evolution with redshift \citep{Lah2007,Delhaize2013,Rhee2013,Rhee2016,Kanekar2016,Rhee2018,Bera2019,Chowdhury2020,Chen2021} and the $M_{\rm HI}$-halo mass relation \citep[][]{Guo2020,Chauhan2021}, among others.  Spectral line stacking has been successfully applied to other spectral lines as well \citep[e.g.][]{Decarli2018,Bischetti2019,Fujimoto2019,Stanley2019,Jolly2020,Jolly2021,Romano2022}.

Despite the fact that spectral line stacking is conceptually simple, it involves several technical and instrumental aspects which can have significant effects on the final results, if they are not accounted for in an appropriate way. In particular, a widely-employed practice is to assign weights to the co-added spectra \citep[e.g. ][see \S\ref{sec:weight}]{Lah2007,Fabello2011a,Delhaize2013,Hu2019,Bischetti2019,Stanley2019}, aiming, for example, at giving less prominence to the contribution of spectra characterized by a larger noise level and/or lower signal-to-noise ratio (SNR). Moreover, the flux is not uniformly transmitted across the field-of-view, but follows the \textit{primary beam} response pattern. This results in $\sim 100\%$ transmission at the centre of the pointing area and a severe flux drop-off towards the angular edges of the cube. Such an effect must be corrected in a proper way \citep[e.g. ][see \S\ref{sec:pb}]{Gereb2013}, otherwise a large bias on $M_{\rm HI}$ estimates is likely to affect the results obtained through stacking.  
Finally, galaxy spectra might contain flux contamination by nearby sources, depending on the specific choice of the angular and spectral aperture adopted to obtain them, and the redshift at which the stacking is performed \citep[e.g.][and references therein, see \S\ref{sec:confusion}]{Elson2016,Elson2019}. On the other hand, choosing a too small angular aperture may yield the extraction of cubelets which do not contain all the \HI{} flux emitted by a galaxy. These aspects need to be accounted for as well, in order not to systematically overestimate, or underestimate, the average $M_{\rm HI}$. 

All the mentioned effects combine together in a complex non-linear way and a careful investigation of their usage must be performed. Although many works have inspected them separately, a fully consistent framework has not been developed yet. In this paper, we use mock interferometric data mimicking realistic MeerKAT 21-cm line cubes and explore the best setup to minimize the impact on the aforementioned operations on results of stacking, to pave the way to per cent accuracy in the exploitation of MeerKAT and other SKA pathfinders forthcoming surveys datasets. In particular, we first assess separately the relative stochastic and systematic deviations of stacking results from the true $M_{\rm HI}$ (known by construction from the simulation used to build the mock cubes) due to random noise, primary beam (hereafter PB) correction and weighting schemes, and then study their cumulative effect. 

An additional aspect which can have a significant effect on the results yielded by stacking is the possible continuum over-subtraction due to the polynomial fitting procedure, responsible for a final underestimation of the signal. We do not simulate the continuum in this work and, hence, leave this for future investigations. Moreover, common biases arising from the definition of the galaxy sample are either selection effects making the sample not representative of the global galaxy population, or the loss of a significant mass of \HI{} due to the definition of a flux-limited sample excluding \HI{}-rich low-mass galaxies. These biases however affect directly the physical conclusions concerning the \HI{} content of galaxy population under analysis, and are not intrinsic biases in the stacking procedure, thereby going beyond the scope of this paper.

This work is organized as follows. In \S\ref{sec:stacking_pipeline_general} we introduce our stacking method and briefly describe its salient features. In \S\ref{sec:stacking_operations} we summarize the stacking variables we aim at testing in this paper. In \S\ref{sec:mock_data} we present the mock dataset we use throughout the paper and the way we generate it and \S\ref{sec:results} presents the core results of this work. We conclude in \S\ref{sec:conclusions}.

Where relevant, we assume a spatially-flat ($\Omega_k=0$) $\Lambda$CDM cosmology with $\Omega_m=0.3$, $\Omega_\Lambda=0.7$ and $H_0=67$ km s$^{-1}$ Mpc$^{-1}$.



\section{Stacking procedure} \label{sec:stacking_pipeline_general}

In this section we summarize the stacking procedure we adopt and present the novel symmetrized stacking technique implemented herein.

\subsection{Obtaining and co-adding spectra} \label{sec:stacking_pipeline}

The stacking analysis presented throughout the paper was performed using a standard spectral stacking procedure, summarized in what follows.

Each individually-undetected \HI{} spectrum is obtained with spatial integration over angular coordinates of a cubelet of size $n\times n \times s$ voxels (where $n$ and $s$ are the number of voxels in the angular and spectral directions, respectively) extracted from the full datacube by relying on the corresponding galaxy optical coordinates and spectroscopic redshift measurements. In our methodology, we choose $n$ and $s$ to have angular and spectral aperture of $3\times \sigma_{\rm beam}$\footnote{We verified that this aperture is larger than the H{\scriptsize I} disc size of the mock galaxies at all redshifts (see also Fig. \ref{fig:mhi_vs_z}), according to the \citet{Wang2016} $M_{\rm HI}-D_{\rm HI}$ relation.}
and $[-1000,1000]$ km s$^{-1}$, respectively. In this way, we limit the problem of evaluating the possible systematic flux excess/defect due to the cubelet size only to the assessment of contamination by nearby sources (see \S\ref{sec:confusion}). 

Each spectrum at observed frequency $f_{\rm obs}$ is then de-redshifted to its rest-frame frequency $f_{\rm rf}$ through $f_{\rm rf}=f_{\rm obs}(1+z)$ and converted in units of velocity in the non-relativistic limit $v/c=z$. Furthermore, spectra are resampled to a reference spectral template, to ensure that all the spectra are binned the same manner in the spectral direction. Lastly, all the spectra are co-added together, thereby giving rise to the final stacked spectrum. Stacking can be performed in units of flux, as well as in units of luminosity or units of mass. Throughout the paper we perform stacking of $M_{\rm HI}$ spectra (i.e., we transform all the spectra from units of flux to units of mass, and then co-add them), being $M_{\rm HI}$ computed as \citep[][]{Weiringa1992}
\begin{equation*}
\centering 
    M_{\rm HI}(\nu) = (2.356\times10^5)\,D_{\rm L}^2\, S(\nu)\,(1+z)^{-1} \, {\rm M}_{\odot}\,{\rm km}^{-1} \, {\rm s}
\end{equation*}
where $D_{\rm L}$ is the luminosity distance of the considered galaxy in units Mpc, $S(\nu)$ is the 21-cm spectral flux density in units Jy and $(1+z)^{-1}$ is a correction factor accounting for the flux reduction due to the expansion of the Universe.

The co-added spectrum can then be expressed as 
\begin{equation}
    \braket{M_{\rm HI}(\nu)} = \frac{\sum_{i=0}^{n_{\rm gal}} M_{{\rm HI},i}(\nu) \times \,w_i }{\sum_{i=0}^{n_{\rm gal}} w_i}
\end{equation}
where $n_{\rm gal}$ is the number of co-added spectra and $w_i$ indicates the weight assigned to each source. In the standard unweighted case, $w_i=1$ and $\sum_i w_i=n_{\rm gal}$. We will present a more detailed discussion of weighting schemes in \S\ref{sec:weight}.

For each galaxy spectrum, a reference spectrum containing no source emission is extracted and co-added to other reference spectra. The reference spectrum is obtained from a cubelet, with centre defined adding a fixed angular offset to the centre of the galaxy cubelet in a random direction, and defined over the same spectral range as the galaxy cubelet. The angular offset is conveniently chosen to guarantee that the reference spectrum is extracted close to the galaxy spectrum, although without overlaps. In this way, we build a reference stacked spectrum representing the null hypothesis, used to further prove that any detection obtained with stacking is not due to noise artifacts. We evaluate the noise level by computing the root mean square (rms) of the noisy channels of the stacked spectrum (hereafter $\sigma_{\rm rms}$). This enables the estimation of the SNR, and the assessment of the noise properties of the datacubes under analysis, e.g. to compare the noise rms versus number of stacked galaxies relation to the $1/\sqrt{N}$ theoretical trend, being $N$ the number of co-added spectra.

Furthermore, we apply a suitable PB correction, presented in details in \S\ref{sec:pb}.

\subsection{Error estimation}

The procedure we adopt to estimate uncertainties on stacked spectra consists in \textit{jackknife resampling} (hereafter jackknnife for shortness) \citep{Quenouille1949,Tukey1956}, known to produce a nearly unbiased estimate of mean square errors. Following the jackknife procedure, one galaxy at a time is deleted from the full sample, running the stacker over each subsample. The error is obtained as the variance of the population of $M_{\rm HI}$ estimates obtained using different galaxy subsamples. 

The jackknife method can be summarized as follows: given the mean
\begin{equation*}
    \bar{x}_i=\frac{1}{n-1}\sum_{j=1,j\neq i}^n x_j
\end{equation*}
of the ensemble of values $x_j$ obtained deleting the $i$th element and
\begin{equation*}
 \bar{x}=\frac{1}{n}\sum_{i=1}^n \bar{x}_i   \quad ,
\end{equation*}
the jackknife variance is computed as the variance of the distribution of $\bar{x}_i$:
    \begin{equation*}\label{eq:jk_variance}
    \sigma^2=\frac{n-1}{n}\sum_{i=1}^n (\bar{x}_i - \bar{x})^2    
    \end{equation*}
where $n$ indicates the size of the set of subsamples. 


\subsection{SNR estimation}

We estimate SNR in two different ways:
\begin{itemize}
    \item maximum flux density (peak) to noise:
    \begin{equation}
    \hspace{1cm}
        {\rm{SNR}}_{\rm{peak}}=S_{\rm{max}}/\sigma_{\rm{rms}}
    \end{equation} 
    where $S_{\rm{max}}$ and $\sigma_{\rm{rms}}$ stand for the peak flux (mass, in the case of this paper) density of the stacked spectrum and the noise rms, respectively;
    \item integrated SNR:
    \begin{equation}
    \hspace{1cm}
    {\rm{SNR}}_{\rm{int}}=\frac{\sum_i^{N_{\rm ch}} S_i\,\Delta v}{\sigma_{\rm rms}\,\Delta v \, \sqrt{N_{\rm ch}}}   
    \end{equation}
    where $N_{\rm ch}$, $S_i$ and $\Delta v$ are the number of velocity channels of the spectral template, the flux (mass) density at channel $i$, and the width of velocity channels (assumed to be constant in our framework), respectively.
\end{itemize}

We use ${\rm{SNR}}_{\rm{peak}}$ to quantify the statistical significance of a detected stacked profile, as deviation from the noise baseline. However, ${\rm{SNR}}_{\rm{peak}}$ is a good estimator for the SNR only in the case where the stacked emission line extends over a narrow velocity range and has a relatively regular Gaussian-like profile. In other cases, where the line is broadened by redshift offset and the resulting line shape is more complex, we rely on ${\rm{SNR}}_{\rm{int}}$ to account for these effects and obtain an estimate for the SNR which is sensitive to all the channels where the \HI{} emission is detected, and not just to the peak of the spectrum. We therefore indicate as SNR the ${\rm{SNR}}_{\rm{int}}$ and rather express ${\rm{SNR}}_{\rm{peak}}$ in terms of $\sigma$-significance throughout the paper, unless stated otherwise.



\begin{figure*}
    \centering
    \includegraphics[width=18cm]{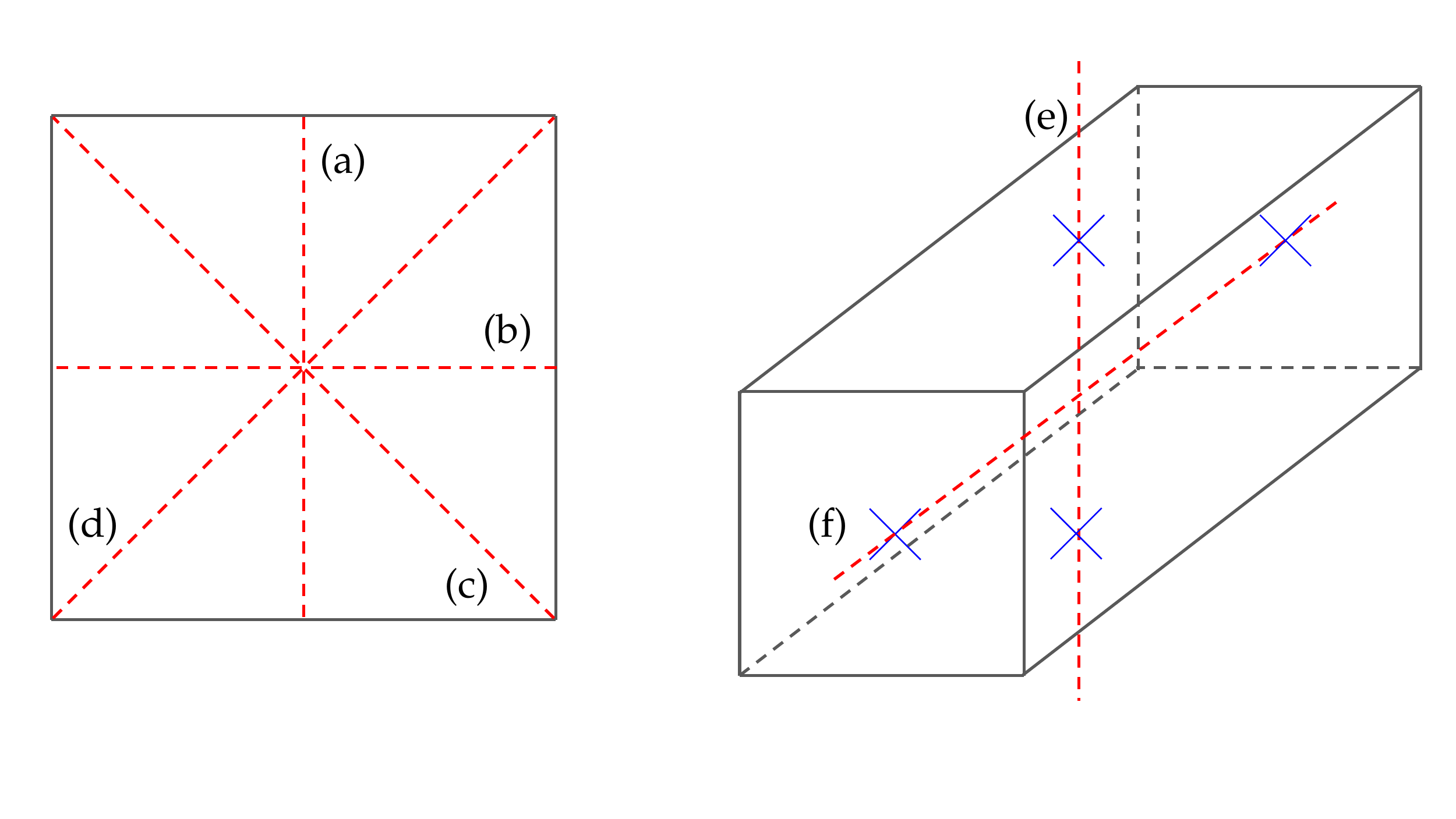}
    \vspace{-1.5cm}
    \caption{A diagram sketching all the symmetry axes of a cubelet. Left: reflection symmetry axes of the square faces on the angular plane. Right: rotation symmetry axes including the spectral dimension.} 
    \vspace{0.5cm}
    \label{fig:symmetry_scheme}
\end{figure*}

\subsection{Spectrum vs cubelet stacking: exploiting cubelet symmetries}\label{sec:symmstacking}

Traditionally, \HI{} galaxy stacking has been performed by co-adding spectra and/or 2D maps. However, \cite{Chen2021} have recently proposed to co-add directly cubelets instead of spectra or images, and to measure the target average $M_{\rm HI}$ through the integration of the final stacked cubelet. This technique has been shown to introduce various advantages over spectral stacking, among which the possibility of deconvolving the stacked cubelet with the stacked point spread function (hereafter PSF), not possible for individual cubelets. If no other operations but stacking are performed (i.e. no deconvolution), spectral and cubelet stacking should yield equivalent results, except for numerical precision errors. 

In this work, we extend the cubelet stacking technique and present a novel idea to fully exploit it to significantly enhance the SNR.
We notice that operating on cubelets implies that one can conveniently take advantage of the geometric symmetries of each cubelet. Considering cubelets with square faces on the angular plane and rectangular faces along the spectral axis, each cubelet can be co-added more than once after applying a symmetry transformation. In fact, each time a cubelet is rotated or symmetrized, a given voxel contributes to the overall stacked cubelet at different positions. In other words, a voxel is co-added to a different voxel of the same cubelet every time a symmetry transformation is applied. Therefore, symmetrized/rotated cubelets can effectively be regarded as independent cubelets from the point of view of SNR statistics (neglecting for a moment the correlation on the scale of the beam on the angular plane, and possible correlations along the spectral axis), although they contain the same underlying \HI{} flux. Therefore, while not altering the final \HI{} signal, cubelet stacking using symmetries (hereafter \textit{symmetrized stacking}) remarkably increases the size of the available galaxy sample without introducing biases. From a different perspective, symmetrized cubelet stacking corresponds to an enhancement of the effective cosmic volume probed by the survey, without altering the mass content of galaxies.

In particular, let us consider a $n\times n\times s$ cubelet, where $n$ and $s$ are the number of voxels in the angular and spectral directions, respectively. From this cubelet (a square cuboid) we can get the following $16$ symmetries:
\begin{itemize}
    \item $90^\circ$ rotation on the angular plane in each channel;
    \item $180^\circ$ rotation on the angular plane in each channel;
    \item $270^\circ$ rotation on the angular plane in each channel;
    \item $2$ reflections around horizontal and vertical axes on the angular plane in each channel;
    \item $2$ reflections around diagonals on the angular plane in each channel;
    \item $180^\circ$ rotation around the zero-velocity axis on the RA-$z$ plane (or on the DEC-$z$ plane, equivalently) + all the aforementioned symmetries.
\end{itemize}

These can be understood more clearly with the aid of the scheme shown in Fig. \ref{fig:symmetry_scheme}, representing on the left symmetry axes on the angular plane and on the right two symmetry axis involving the spectral direction. The possible symmetries are reflections with respect to axes (a), (b), (c) and (d), $4$ rotations around axis (f), and the same after rotating the cubelet by $180^\circ$ around axis (e). 

While symmetries on the angular plane pose no conflict since the same gridding is used throughout the whole (RA,DEC) ranges, in order to be able to apply symmetry transformations after rotating the cubelet around axis (e) a suitable spectral resampling might be necessary, in order to ensure that all bins in the spectral direction have the same width. 

This results in $16$ possible symmetry transformations, implying a potential increase of the galaxy sample size by a factor $16$ and, hence, of the the SNR by a factor $\sqrt{16}=4$. However, the angular or spectral integration performed to produce stacked spectra or images, respectively, make some of the symmetries vanish. In particular, a stacked spectrum is sensitive only to $180^\circ$ rotation around axis (e), while a stacked image is sensitive only to reflections and rotations occurring on the angular plane. As a result, symmetrized stacking effectively produces $2$ symmetry transformations in the case of a stacked spectrum and $8$ symmetry transformations in the case of a stacked image.

We notice that in our specific case, applying symmetrized stacking to a stacked spectrum is equivalent to perform a reflection symmetry directly on spectra with respect to the $v=0$ channel and thereby implies an increase of the galaxy sample size by a factor $2$ and a potential SNR gain by a factor $\sqrt{2}\sim 1.4$.

Moreover, as anticipated, voxels are correlated on the sky on the scale of the beam, and hence cubelet symmetrization may not yield the expected noise level drop and SNR gain when symmetries on the angular plane remain effective (e.g. in the stacking of images). In particular, we expect this to hold true especially if the stacked sources are unresolved, or their size is barely larger than the beam size. In such a case, one may not get the expected SNR gain, but rather experience a reduced efficiency of the symmetrization technique. Similarly, correlations along the spectral axis may arise depending on channels size and velocity resolution. In this work, we are interested in testing the stacking procedure based on spectra, so the cubelet symmetrization technique eventually results in being equivalent to a simple flipping of the spectra of single galaxies. We test this technique to produce stacked spectra in \S\ref{sec:symmetry}, and leave the investigation of the results when applied to stacked images for later work.


\section{Instrumental and technical variables in stacking} \label{sec:stacking_operations}

In this section we introduce the technical and instrumental aspects which need to be taken into account and corrected when stacking \HI{} galaxy spectra, and which we test in \S\ref{sec:results}.

\subsection{Primary beam}
\label{sec:pb} 
In general, the sensitivity of a radio-telescope across the footprint is non-uniform and follows the normalized PB response pattern $f\equiv S(\rho)/S(\rho=0)$, where the $\rho$ coordinate indicates the pointing offset and $\rho=0$ denotes the centre of the pointing. For the MeerKAT radiotelescope, \cite{Mauch2020} found that the primary beam is well matched by the attenuation pattern resulting from cosine-squared power illumination \citep[][]{Condon2016}. In this work, we rather adopt a simplistic model and assume the Gaussian approximation, i.e. that the PB follows a 2-dimensional Gaussian function
$f(\rho) = \exp\left(-0.5\times(\rho/\theta)^2\right)$, where $\rho$ denotes the pointing offset (see above) and $\theta$ the standard deviation.

In the case of detected emission in a voxel $i$ of a datacube, the primary beam correction is obtained by simply dividing the observed flux $S_{21,\rm{obs}}(i)$ by $f(i)$. I.e., the true flux is obtained as $S_{21,\rm{true}}(i)=S_{21,\rm{obs}}(i)/f(i)$.

Conversely, in usual stacking applications the \HI{} emission is undetected and applying the aforementioned PB correction would imply boosting the noise and the underlying signal by the same factor, thus obtaining no SNR gain. Under these circumstances, a spectrum extracted at a position where $f\ll1$ will end up dominating the whole stack, which would then result in a very noisy spectrum, potentially with no detected emission.

To correct for PB flux attenuation in our stacking pipeline, we adopt the following scheme \citep[e.g.,][]{Gereb2013,Gereb2015, Hu2019}:
\begin{equation} \label{eq:pb_raw}
S(\nu)=\frac{\sum_i f_i S_i(\nu)}{\sum_i f_i^2}    
\end{equation}

where the correction is obtained through a weighted average of the quantity $S_i/f_i$ by its average $f_i^2$ (defined over the voxels of the cubelet), where the weights $f_i^2$ enable to obtain the sought enhancement of the stacked signal.



We notice that the PB correction does not represent a point of concern when the galaxy sample is constituted by galaxies located in angular areas of the cube where $f\sim 1$. This happens when e.g. single pointings to target galaxies are performed, ensuring that they are found close to the centre of the field of-view, or when mosaics of single partially-ovelapping pointings are realized, thereby compensating the flux attenuation. Instead, the primary beam flux reduction is a major issue when single pointings are performed and the galaxy sample is selected to include also sources in angular regions where $f\ll1$. Yet, the problem persists also in mosaic observations with small area of overlap between different pointings. 

In the view of the future SKA project, a robust treatment and assessment of PB correction is needed. Although the employment of Eq. (\ref{eq:pb_raw}) is well-justified by mathematical and physical arguments and should in principle yield unbiased $M_{\rm HI}$ estimates (see Appendix \ref{sec:appendix}), a systematic assessment of its accuracy in the presence of noise and under realistic observational conditions is still lacking. In addition, Eq. (\ref{eq:pb_raw}) is often used in combination with weighting schemes (see \S\ref{sec:weight}), making an eventual bias in $M_{\rm HI}$ estimate with stacking a degenerate effect between PB correction and the adopted weighting scheme.

\subsection{Weighting schemes} \label{sec:weight}

In addition to the basic (unweighted) stacking procedure, it has become a customary approach to assign a weight to each co-added spectrum to optimize SNR and accuracy, depending on the properties of the underlying noise and on other galaxy properties, such as distance. The goal of these further refinements is to suppress unwanted boosted noise features (e.g. spikes, artifacts, residuals of continuum), which may potentially endanger the overall procedure. To this end, it seems a natural choice to weight spectra by the inverse of an estimator of the noise level in each galaxy spectrum. Following this line of reasoning, \cite{Lah2007} and \cite{Fabello2011a} proposed $w_i=1/\sigma_{i,\rm{rms}}$ (L07 hereafter) and $w_i=1/\sigma_{i,\rm{rms}}^2$ (F11 hereafter) , being $\sigma_{i,\rm{rms}}$ the noise rms of the $i$th galaxy spectrum. Building on such works, \cite{Delhaize2013} considered the possibility of implementing the scheme $w_i=1/(\sigma_{i,\rm{rms}}\,d_l^2)^2$, arguing that such a scheme would optimize even further the SNR, at the expense of a larger cosmic variance effect. Lately, \cite{Hu2019} generalized this approach as $w_i=1/(\sigma_{i,{\rm{rms}}}^2\,d_l^\gamma)$, studying the results in terms of $M_{\rm HI}$ and SNR as a function of $\gamma$. In this paper we test the case $\gamma=4$ and refer to it as H19, even though such a specific scheme has first been proposed by \cite{Delhaize2013}, to account for the fact that it represents just one choice among different possible distance-based weightings.


As for Eq. (\ref{eq:pb_raw}), no systematic checks of the accuracy of the different weighting schemes has been performed in the aforementioned works, due to the intrinsic lack of knowledge about the underlying true $M_{\rm HI}$ distribution of galaxies. In fact, assessing the accuracy of these techniques requires a detailed modelling of simulated datacubes, including realistic noise, PSF and PB.

Assuming that the $i$th galaxy is assigned a weight $w_i$ and has average PB response $f_i$, the final stacked flux can be written as:
\begin{equation} \label{eq:pb_weight}
S(\nu)=\frac{\sum_i f_i\,w_i\, S_i(\nu)}{\sum_i f_i^2\,w_i}    
\end{equation}

We notice that PB correction and weighting schemes combine in a non-linear fashion, with no a-priori guarantee that the final stacked signal is a faithful proxy for the real underlying \HI{} signal. In what follows, we address this question and test our method in a systematic manner. As anticipated, we evaluate the accuracy of Eq. (\ref{eq:pb_weight}) using mock observations constructed as explained in \S\ref{sec:mock_data}, where the \HI{} galaxy masses are available by construction.

\subsection{Source confusion} \label{sec:confusion}

A long-standing systematics of spectral stacking consists in flux contamination by nearby sources, due to the intrinsic proximity of galaxies in space and the finite resolution of the telescope. This issue, also referred to as \textit{source confusion}, has been tackled in different ways, either by means of analytical arguments \citep[e.g.][]{Fabello2012a,Delhaize2013,Jones2016,Hu2019} or through an assessment based on simulated datacubes \citep{Elson2016,Elson2019}. Having in this work simulated cubes directly available (see \S\ref{sec:mock_data}), we adopt the second strategy.
At the probed distances, and with the chosen beam FWHM $\sigma_{\rm beam}$ (see \S\ref{sec:mock_data}) and $3\times\sigma_{\rm beam}$ angular aperture, we find source confusion to overestimate by $\lesssim 2\%$ the final $\braket{M_{\rm HI}}$ and therefore regard it as a negligible effect. 
We point out that our approach is particularly suitable for future applications of stacking a $z>0.1$, where the contribution of confusion becomes relevant \citep{Elson2016,Elson2019}.  



\begin{figure*}
    \centering
    \includegraphics[width=18cm]{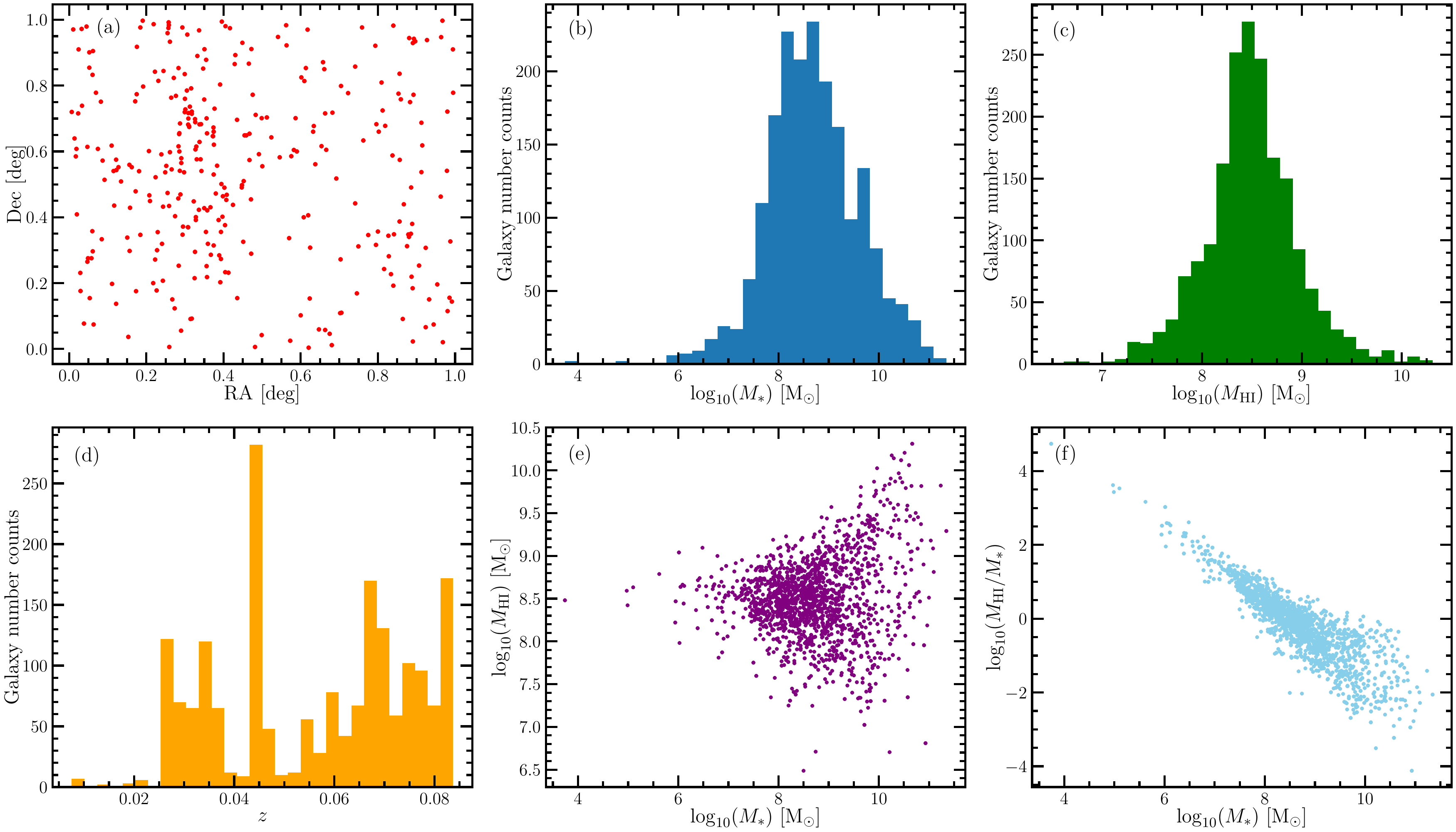}
    \caption{Properties of the modelled galaxies injected in the mock datacubes presented in \S\ref{sec:mock_data}. Panel (a): galaxy positions projected on the angular plane (one cube). Panel (b): $\log_{10}(M_*)$ galaxy number counts distribution (all cubes). Panel (c): $\log_{10}(M_{\rm HI})$ galaxy number counts distribution (all cubes). Panel (d): $z$ galaxy number counts distribution (all cubes). Panel (e): $\log_{10}(M_{\rm HI})$ as a function of $\log_{10}(M_*)$ (all cubes). Panel (f): $\log_{10}(M_{\rm HI}/M_*)$ as a function of $\log_{10}(M_*)$ (all cubes)}. 
    \label{fig:galaxy_props}
\end{figure*}

\begin{table}
    \centering
    \begin{tabular}{ll}
    \toprule
    \toprule
    \textbf{Mock data} & \\
    \toprule
    \textbf{Survey parameter} & \textbf{Value}\\
    \midrule
    Total area     & $6\times 1$ deg$^2$\\
    Number of images     & $6$\\
    Frequency resolution & $209$ kHz\\
    Velocity resolution & $44$ km s$^{-1}$\\
    Frequency range & $1.310-1.420$ GHz \\
    Velocity range & $86-24205$ km s$^{-1}$ \\
    Beam (FWHM) &  $12.0^{\prime\prime}\times12.0^{\prime\prime}$ \\
    Pixel size     & $4.0^{\prime\prime}$ \\
    Image size & $900\times900$ pixels \\
    Total number of galaxies & $1901$\\
    Number of \HI{}-undetected galaxies & $709$\\
    \bottomrule
    \bottomrule
    \end{tabular}
    \caption{Summary of the details of mock data presented in \S\ref{sec:mock_data} and used in this paper.}
    \label{tab:mock_data}
\end{table}

\section{Mock dataset}\label{sec:mock_data}

The simulated datacubes have been generated according to the \cite{Obreschkow2014} flux-limited mock galaxy catalogue, based on the SKA Simulated Skies semi-analytic simulations (S$^3$-SAX) and therefore on the physical models described in \cite{Obreschkow2009a,Obreschkow2009b,Obreschkow2009c}. The catalogue spans a $\sim10^2\,\rm{deg}^2$ area in the sky and presents detailed \HI{} and optical properties for millions of galaxies at $0<z<1.2$, including realistic masses and sizes of \HI{} discs, and positions. In particular, we take advantage of the latter to ensure that our simulated mock galaxies are distributed in space with credible clustering properties, an aspect of major importance for what concerns source confusion (see \S\ref{sec:confusion}).  

The spatial and spectral properties of the \HI{} emission for each galaxy are then realistically modelled in three dimensions, through an appropriate parametrization of rotation curve and \HI{} radial density profile \citep{Elson2016,Elson2019}. In particular, we parametrize the azimuthally-averaged radial distribution of $M_{\rm HI}$ density as
\begin{equation} \label{eq:radial_profile}
\Sigma_{\rm HI}(r) = \frac{A\,\exp{(-r^2/(2h^2))}}{1+\beta\,\exp{(-1.6\,r^2/(2h^2))}} 
\end{equation}
where $A$ is a normalization parameter used to match the total $M_{\rm HI}$, $h$ is the standard deviation of the Gaussian and is chosen to be $h=R_h$, where $R_h$ is the disc scale length from \cite{Obreschkow2014}, $r$ is the distance from the centre of the galaxy, and $\beta$ regulates the central \HI{} concentration, yielding a central \HI{} depression if $\beta>0$.  

On the other hand, rotation curves are modelled using a Polyex profile \citep{Giovanelli2002}:
\begin{equation} \label{eq:rotation_curve}
    v(r) = v_0 \, \left[ 1- \exp{(-r/r_0)}\right] \, \left(1+\frac{\alpha\,r}{r_0}\right)
\end{equation}

where $r$ is the distance from the centre of the galaxy, $v_0$, $r_0$ and $\alpha$ are free parameters, for which we adopt values from \cite{Catinella2006} \citep[see Table 1 of ][for a summary]{Elson2016}, based on the galaxy $I$-band magnitude. In this way, synthetic galaxies will have modelled rotation curves based on empirical measurements.

With physical properties expressed in Eq. (\ref{eq:radial_profile}) and (\ref{eq:rotation_curve}), a galaxy is then modelled as a collection of \HI{} clouds, under thin disc and axisymmetry approximations.

All the galaxies with non-zero $M_{\rm HI}$ from the \cite{Obreschkow2014} catalogue which fall within the chosen spatial and spectral ranges are therefore modelled as described above and are interpolated on a regular grid with suitable pixel size ($4^{\prime\prime}$ in our case) at the positions defined in the catalogue. This constitutes a preliminary noise-free version of the cube. Optionally, the resulting cube is multiplied by a model for the normalized primary beam (see below and \S\ref{sec:pb}). In a second stage, Gaussian noise (completely specified by its mean and variance) is assigned to the each cell of the datacube through random sampling, with standard deviation chosen to match the desired noise level. Lastly, the resulting mesh is spatially smoothed with an appropriate synthetized beam model. 

We generate mock data corresponding to $6$ different telescope pointings, one cube per pointing, spanning an area of $6\,\rm{deg}^2$ ($1\,\rm{deg}^2$ each) and a frequency range $1310<\nu<1420\,\rm{MHz}$ ($0.005<z<0.084$ for 21-cm emission).  Each cube has been extracted from the simulation at a position different to the one of the other cubes, with no overlapping regions. The parameters of our resulting dataset are designed to roughly match the features of the MIGHTEE-\HI{} Early Science data \citep{Maddox2021}, and therefore allow us to perform our analysis under a realistic observational scenario. Furthermore, creating many smaller datacubes instead of single cube with very large angular footprint ensures to have sufficient statistics in the $f\sim1$ primary beam regions and to alleviate cosmic variance due to the particular choice of the mock-surveyed area of the sky.

Fig. \ref{fig:galaxy_props} illustrates some key features of the galaxies used to generate the mock data. Panel (a) shows the projection of galaxy positions on the angular plane in one cube. Panels (b), (c) and (d) represent the distributions of $\log_{10}(M_*)$, $\log_{10}(M_{\rm HI})$ and $z$, respectively, of the entire dataset. Panels (e) and (f) show $\log_{10}(M_{\rm HI})$ and $\log_{10}(M_{\rm HI}/M_*)$, respectively, as a function of $\log_{10}(M_*)$, for all the galaxies in our sample. 
As expected from the \citet{Obreschkow2014} simulation, galaxies display realistic clustering properties (panels a and d). In particular, we notice that the redshift distribution is not uniform, consistently with the succession of overdense and underdense cosmic structures. The presence of an overabundance in galaxy number counts at $z\sim 0.045$ is consistent with the presence of a galaxy cluster in the mock-observed sky area.

Table \ref{tab:mock_data} summarizes the details of our simulated data.
In order to be able to perform a meaningful comparison with realistic observational data, we design our synthetic cubes to match the features of data acquired with the MeerKAT radio telescope. Noise is randomly-sampled according to a Gaussian distribution with zero-mean and standard deviation $\sigma_{\rm n}\sim 4\times 10^{-5}\,\rm{Jy}\,\rm{beam}^{-1}$. We neglect channel-dependent fluctuations of the noise rms, as well as intrinsic noise rms variations with frequency and radio frequency interference phenomena. We notice that neglecting these aspects implies devising an optimistic model for the noise, which may have a beneficial impact on stacking results. However, we are here interested in studying the performance of stacking under the ideal assumption that noise follows Gaussian statistics. In fact, modelling additional noise features would again bring in new potential degeneracies on the estimation of the contribution of different stacking variables to the deviation of final results from the ground truth. In this sense, we regard our model for the noise as suited to accomplishing our goals. Both the synthetized beam and the primary beam are modelled as two-dimensional rotationally-symmetric Gaussian functions, of standard deviation $\sigma_{\rm beam}=12^{\prime\prime}$ and $\sigma_{\rm PB}=10^{\prime}$, respectively. Although we are well aware that such a choice represents an approximation to the true geometry of these objects \citep[see \S\ref{sec:pb} for a summary on the MeerKAT PB and e.g.,][for typical deviations of the MeerKAT beam from rotational symmetry]{Ponomareva2021}, for practical purposes we regard the mismatch between our model and the true shapes to be negligible, as long as the same model is consistently used in all the synthetic cubes we create.

\begin{figure*}
    \centering
    \includegraphics[width=18cm]{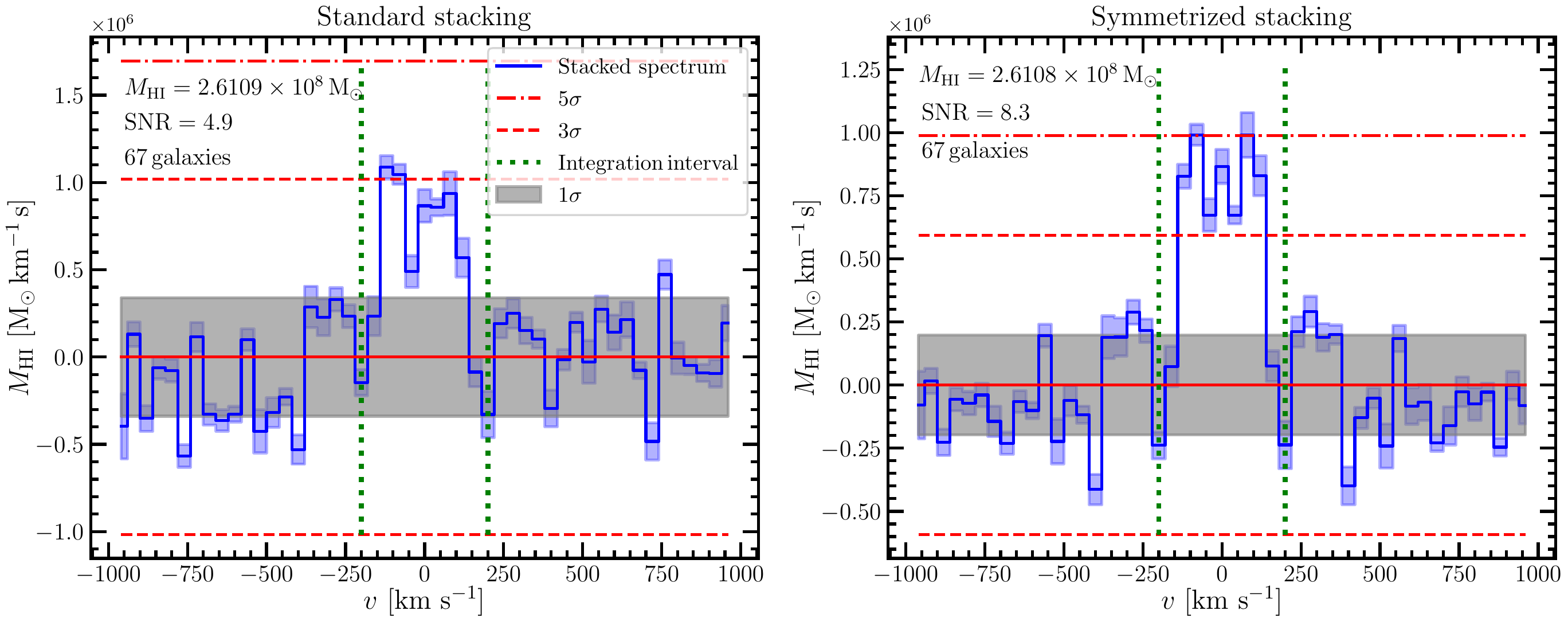}
    \caption{Comparison between stacked spectra resulting from standard spectral stacking (left) and symmetrized stacking (right, equivalent to effectively using two different spectra per galaxy, see \S\ref{sec:symmstacking}). The symmetrized stacking is shown to reproduce the $\braket{M_{\rm HI}}$ yielded by spectral stacking, with negligible $\Delta M_{\rm HI}\ll 1\%$ deviation. The SNR of the resulting detected $\braket{M_{\rm HI}}$ signal is enhanced by a factor $\sim 1.7$, larger but compatible within uncertainties with the theoretical SNR gain $\sim1.4$ expectation. Blue lines indicate the resulting spectra, with corresponding shaded areas representing uncertainties obtained through jackknife resampling. Gray shaded areas, red dashed and red dashed-dotted lines stand for $1\sigma$, $3\sigma$ and $5\sigma$ noise rms levels.} 
    \label{fig:symmetry}
\end{figure*}

\begin{figure*}
    \centering
    \includegraphics[width=18cm]{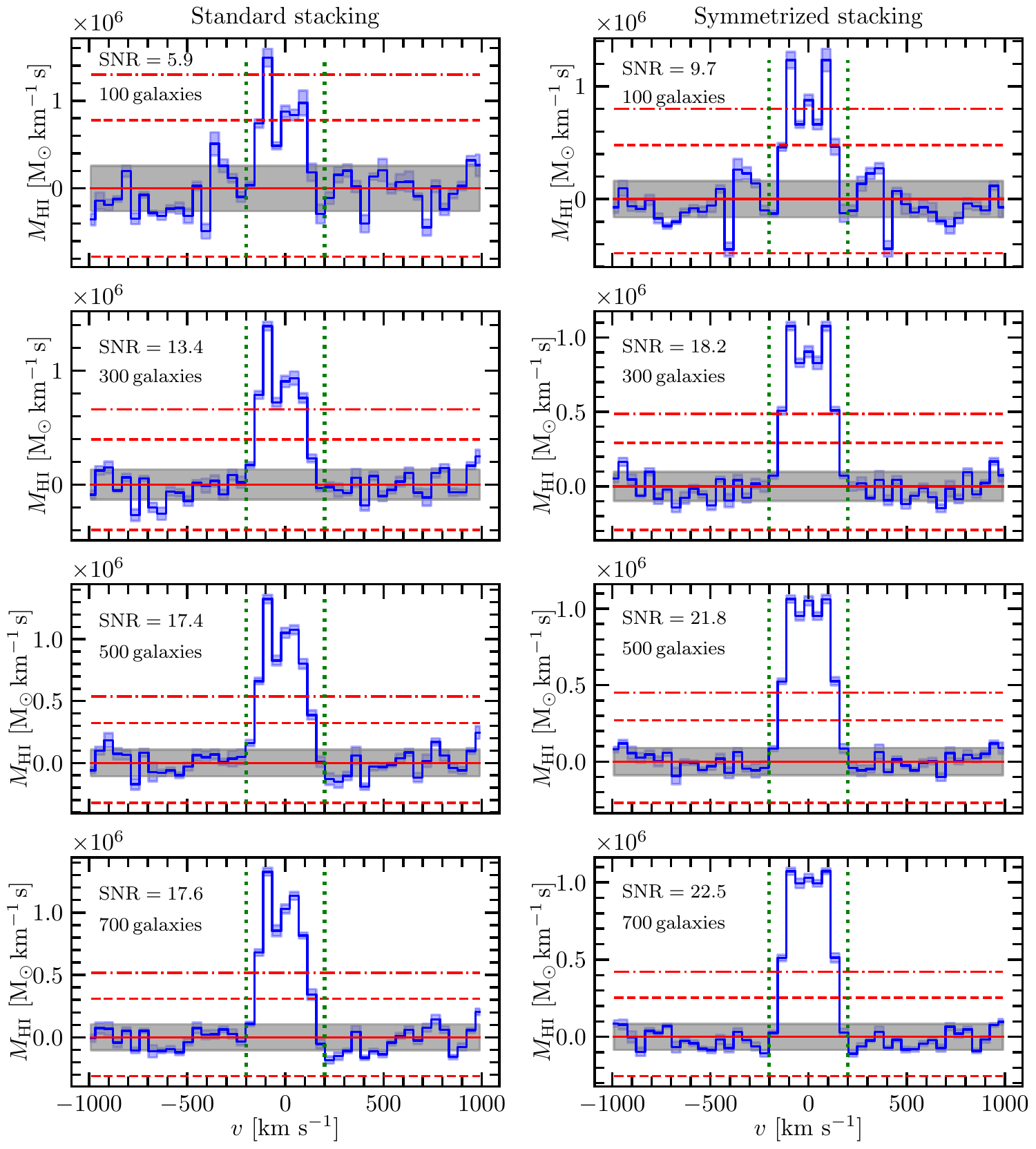}
    \caption{Comparison between stacked spectra resulting from standard spectral stacking (left column) and symmetrized stacking (right column, equivalent to effectively using two different spectra per galaxy, see \S\ref{sec:symmstacking}), as a function of the number of stacked galaxies. The symmetrized stacking is shown in all the cases to enhance the SNR of the $\braket{M_{\rm HI}}$ signal yielded by standard stacking by a factor compatible within uncertainties with the $\sim1.4$ SNR gain expectation. Blue lines indicate the resulting spectra, with corresponding shaded areas representing uncertainties obtained through jackknife resampling. Gray shaded areas, red dashed and red dashed-dotted lines stand for $1\sigma$, $3\sigma$ and $5\sigma$ noise rms levels.} 
    \label{fig:symmetry_numgal}
\end{figure*}

In our framework, each pointing comes with four associated cubes: 
\begin{itemize}
    \item without primary beam effect (\textit{flat}, hereafter) noise-free cube;
    \item with primary beam effect included (\textit{PB uncorrected}, hereafter) noise-free cube;
    \item flat noise-filled cube;
    \item PB uncorrected noise-filled cube.
\end{itemize}

Among these, only the PB uncorrected noise-filled cube is equipped with all the features of real observational data. The full dataset allows us to perform a direct assessment of the joint degenerate impact of noise, PB correction and weighting schemes. 

The total sample of galaxies consists of 1901 sources, of which 709 are \HI{}-undetected according to a $3\sigma_{\rm n}$ flux cut. In particular, we have extracted $9\times9\times 41$ (RA, DEC, $z$) voxels cubelets and flagged as detections all the sources containing more than $12$ voxels with flux $>3\sigma_n$ (corresponding to $>0.3\%$ of the total number of voxels per cubelet, i.e.  outside the $3\sigma$ confidence interval). We consider only undetected sources in our stacking experiments, to mimic realistic observational conditions and not to strongly bias the stacking results. 

Furthermore, to consistently compare results obtained using flat and PB uncorrected cubes, we define the detections catalogue upon the flat cube. In fact, some galaxies might be undetected in the PB uncorrected cubes as an effect of the PB flux attenuation, but detected in the corresponding flat cubes.


\section{Results and discussion} \label{sec:results}

In this section we present the results of our analysis. Throughout the section, we make use of all the $6$ sets of mock cubes described in \S\ref{sec:mock_data}. We first assess the accuracy of symmetrized stacking. Then, we investigate the accuracy of the aforementioned stacking operations using the exact $z$ values from the simulation. Later, we add a suitable redshift redshift $\Delta z$ to galaxies and repeat the procedure using a new list of redshifts $z'=z+\Delta z$, to mimic the redshift uncertainty characterizing real observations.  

\begin{figure*}
    \centering
    \includegraphics[width=18cm]{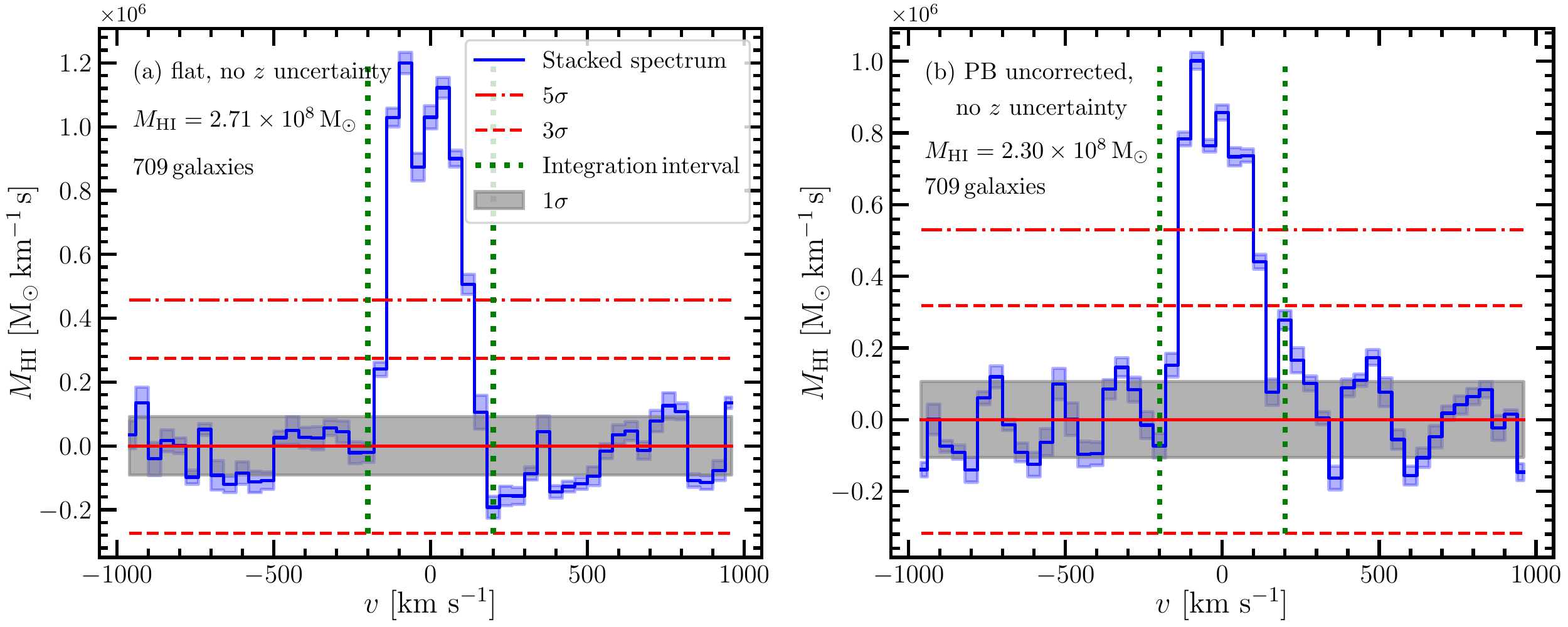}
    \caption{Stacked spectra obtained including all the available undetected sources in (a) flat cubes without redshift uncertainty, and (b) PB uncorrected cubes without redshift uncertainty. Blue lines indicate the resulting spectra, with corresponding shaded areas representing uncertainties obtained through jackknife resampling. Gray shaded areas, red dashed and red dashed-dotted lines stand for $1\sigma$, $3\sigma$ and $5\sigma$ noise rms levels.} 
    \label{fig:without_z_offset}
\end{figure*}

\subsection{Symmetrized stacking}
\label{sec:symmetry}

Fig.~\ref{fig:symmetry} shows the results of standard spectral stacking in left panel and obtained through symmetrized stacking in right panel. The results evidence the statistical significance of the detection is boosted from $\sim3.2\,\sigma$ to $\sim 5\,\sigma$ ($\sim 1.8\,\sigma$ difference). When comparing integrated SNR, spectral stacking yields $\rm{SNR}\sim 4.97\pm 0.33$, while symmetrized cubelet stacking yields $\rm{ SNR}\sim 8.31\pm 0.21$, thereby enhancing the SNR by a factor $\sim 1.67\pm 0.39$, larger but compatible within uncertainties with the theoretical SNR gain $\sim1.41$ expectation. The $\braket{M_{\rm HI}}$ estimate, obtained integrating the resulting spectra over the $[-200,200]\,\rm{km s}^{-1}$ velocity range (enclosed within the green vertical dotted lines), is matched by the two techniques with deviation $\Delta M_{\rm HI}\ll1\%$, i.e. with extremely high degree of consistency. We also notice that the resulting stacked spectrum after symmetrization is, as expected, symmetric with respect to the $v=0$ symmetry axis.

Fig. \ref{fig:symmetry_numgal} shows a comparison between the results of standard spectral stacking (left column) and obtained through symmetrized stacking (right column), as a function of the number of stacked galaxies (different rows). In all the studied cases, the $\braket{M_{\rm HI}}$ is again matched by the two techniques with $\ll 1\%$ deviations, and the SNR gain achieved when using symmetrized stacking over standard stacking is compatible within uncertainties with the $\sim 1.4$ gain theoretical expectation.  

While this technique adds little when robust detections are obtained with spectral or (non-symmetrized) cubelet stacking, it turns out to be in principle very useful in the case of detections with low SNR. In fact, symmetrized stacking has the potential of turning a weak detection, or even a non-detection, into a more solid detection.

We therefore point out that this technique, introduced for the first time (to the knowledge of the authors) in this paper, provides a simple and efficient way of gaining significance and robustness on the stacked signal, without introducing any bias in the final estimate of $\braket{M_{\rm HI}}$. 

In this paper, we employ this technique whenever the galaxy sample becomes too small to yield a robust detected $\braket{M_{\rm HI}}$ signal, as sometimes happens when e.g. excluding galaxies located in angular regions where $f<0.55-0.6$. In particular, we adopt symmetrized stacking when SNR$<5$.

\subsection{Stacking without redshift uncertainty}

We start by considering galaxies at their original redshift $z$, available by construction from the simulation.

Fig. \ref{fig:without_z_offset} shows the result of stacking applied to galaxies without $z$ offset, in the flat (panel a) and PB uncorrected (panel b) cubes. In (a), the resulting stacked emission line has a clear double-horn profile, which is not very common to observe in realistic stacking studies based on observations. In (b), the double-horn profile is not visible as clearly as in (a), although there are clues pointing towards a shape with two peaks. It is worth noticing that the stacked emission line extends out to $\pm 200$ km s$^{-1}$, narrower than what is typically found in stacking based on observational data ($> 250-300$ km s$^{-1}$).

\begin{figure*}
    \centering
    \includegraphics[width=18cm]{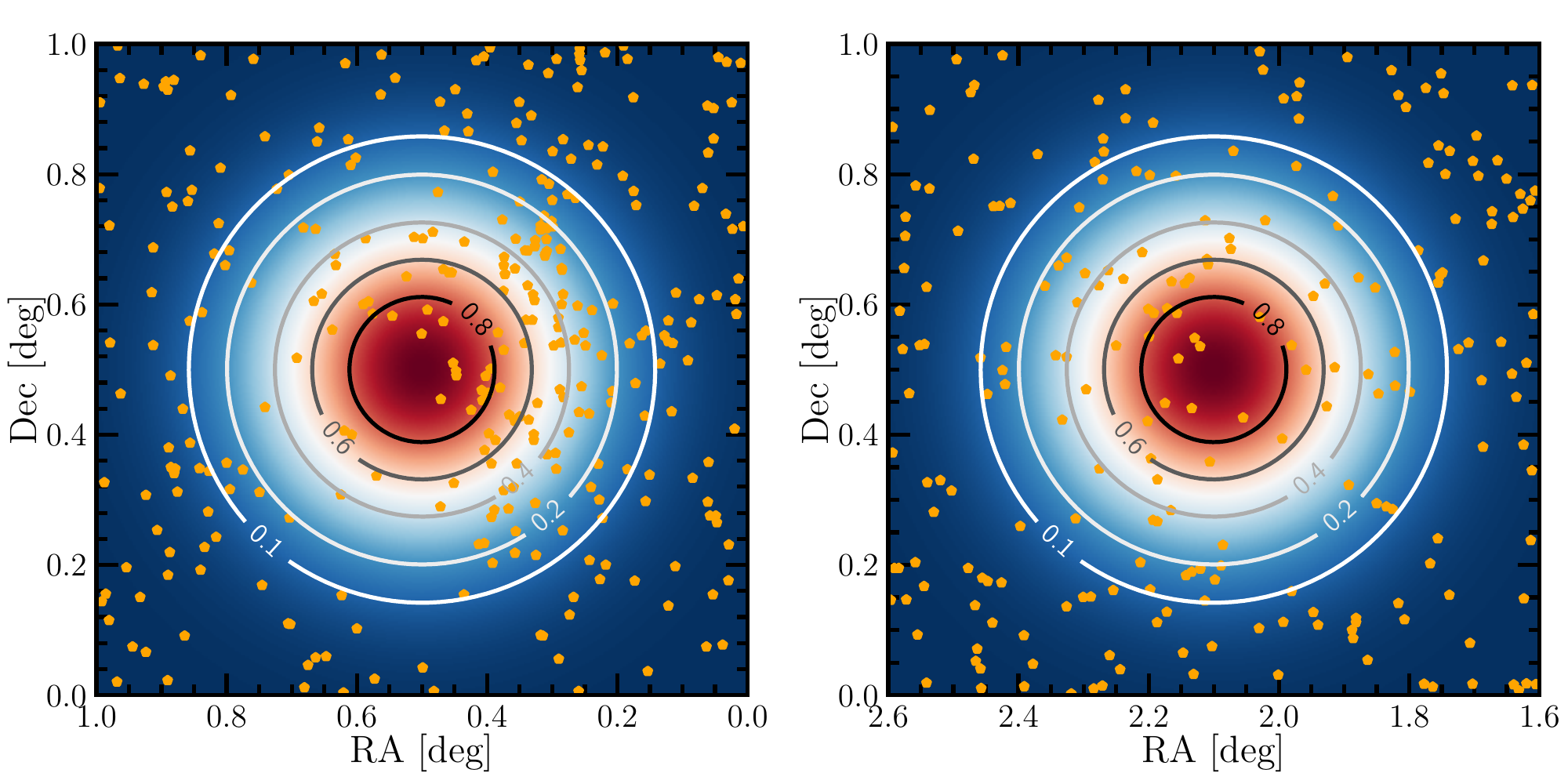}
    \caption{Projection of \HI{}-undetected galaxies positions on the angular plane in two different cubes. The background is color-coded as function of the normalized primary beam (blue to red from 0 to 1). Contours indicate the $f$ levels indicated  by their labels.} 
    \label{fig:footprint}
\end{figure*}

\begin{figure*}
    \centering
    \includegraphics[width=18cm]{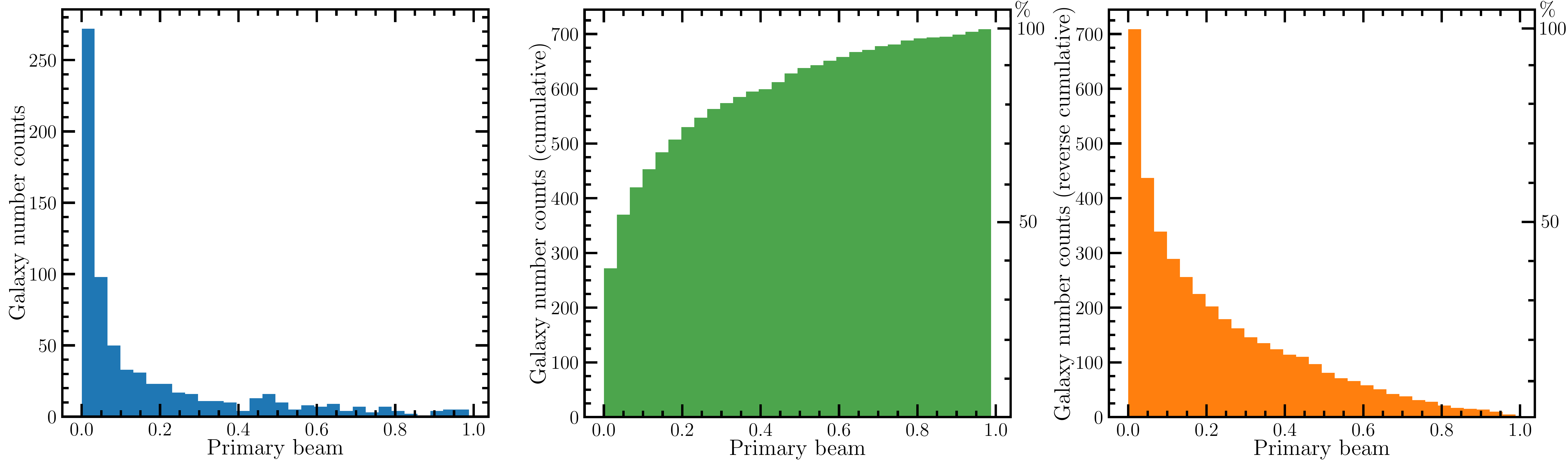}
    \caption{Left: mock galaxy number counts distribution as a function of PB. Centre: cumulative mock galaxy number counts distribution as a function PB. Right: reverse cumulative mock galaxy number counts distribution as a function PB. Only \HI{}-undetected galaxies were considered to generate the three panels of this plot.} 
    \label{fig:pb_distribution}
\end{figure*}

The core of our analysis consists in comparing the $\braket{M_{\rm HI}}$ values obtained in the flat noise-free cubes results -- representing the true $\braket{M_{\rm HI}}$ which we seek to recover with stacking (plus source confusion, which can be subtracted a posteriori) -- with stacking results obtained in:
\begin{itemize}
\item the PB uncorrected noise-free cubes, to assess the accuracy of the PB correction alone;
\item the flat noise-filled cubes, to assess the accuracy and precision of stacking in the presence of noise alone;
\item the (realistic) PB uncorrected noise-filled cubes, to jointly assess the impact of the PB correction and of noise.
\end{itemize}
To evaluate the accuracy with which we recover our ground truth $\braket{M_{\rm HI}}$, we use as metric the following percentage mass residuals:
\begin{equation*}
    \Delta M_{\rm HI}=100\%\times(\braket{M_{\rm HI}}/\braket{M_{\rm HI}}_{\rm true}-1)
\end{equation*}
which is sensitive both to the amplitude and to the sign of a given deviation. 

Moreover, we exclude by our sample all galaxies below a given threshold PB$_{\rm th}$, starting from PB$_{\rm th}=0$ and gradually highering it, and study the trend of $\Delta M_{\rm HI}$ as a function of PB$_{\rm th}$. The rationale behind this is that we aim to find the best trade-off between the choice of PB$_{\rm th}$, the resulting SNR and the potential $M_{\rm HI}$ bias, and to devise a method to perform statistical corrections a posteriori. In fact, increasing PB$_{\rm th}$ typically means reducing significantly the number of sources available to stack, and thus the potential SNR gain. On the other hand, \HI{}-undetected sources in regions where $f\ll1$ contribute little signal and mostly noise to the stack, at risk of worsening the quality of the final result. Therefore, as anticipated one would like to find which choice of PB$_{\rm th}$ ensures a safe SNR and to know how to correct an eventual bias in $M_{\rm HI}$ estimation. 

\begin{figure*}
    \centering
    \includegraphics[width=18cm]{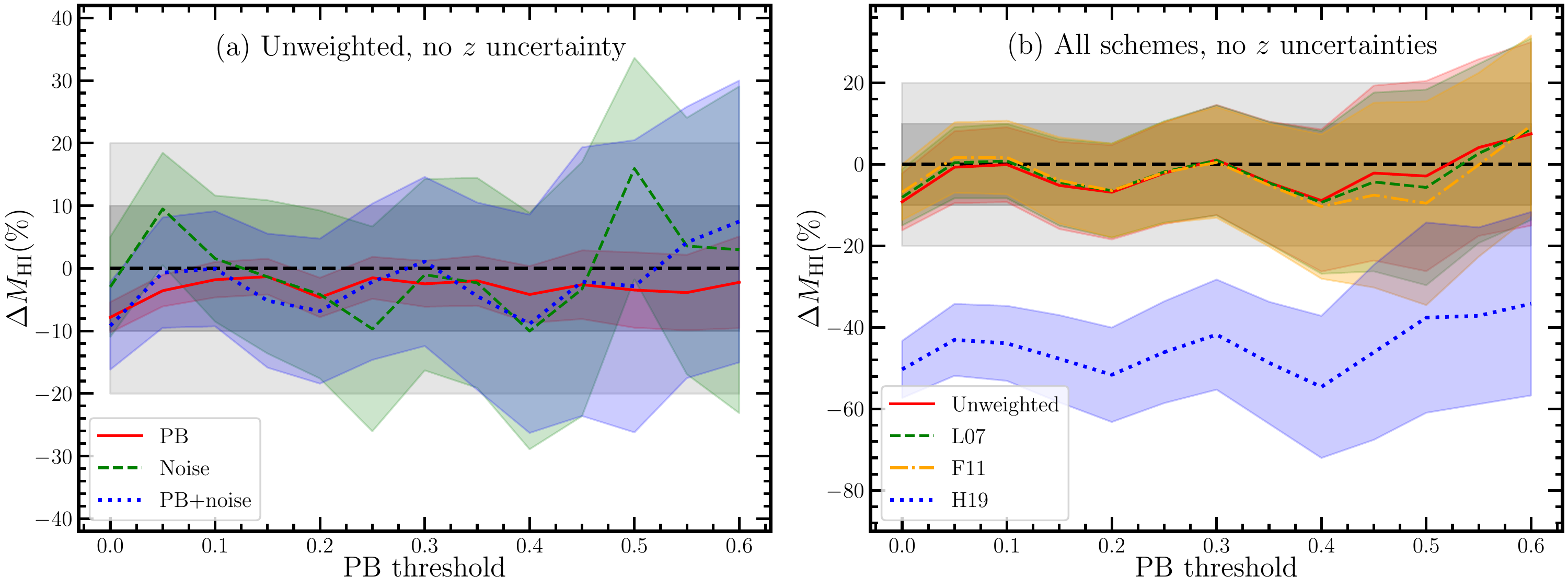}
    \caption{Percentage mass deviation $\Delta M_{\rm HI}$ as a function of PB$_{\rm th}$. Panel (a) shows the results of stacking in PB uncorrected noise-free cubes (red solid line), in flat noise-filled cubes (green dashed line) and in PB uncorrected cubes noise-filled (blue dotted line), representing the effects of PB correction alone, noise alone, and the combination of the two, respectively. No redshift offset and no weighting schemes (unweighted case) are applied. Panel (b) shows the results of stacking in PB uncorrected cubes noise-filled (applying PB correction in the presence of noise), with $4$ difference weightings: unweighted (red solid), L07 (green dashed), F11 (orange dashed-dotted), H19 (blue dotted). No redshift offset is applied. Shaded areas represent $1\sigma$ uncertainties around the curves they are associated to.} 
    \label{fig:deltamass_wo_z_offset}
\end{figure*}

\subsubsection{Spatial distribution of galaxies within the footprint}

To understand which is the relative spatial distribution of galaxies with respect to the underlying primary beam, we start with a visual inspection of the positions of galaxies projected onto the angular plane for two different simulated datacubes, as shown in Fig. \ref{fig:footprint}. Here, the background is color-coded as a function of $f$ (growing $f$ from blue to red), contours indicate different $f$ levels, and the positions of galaxies are represented as orange symbols. It turns out rather clearly that the vast majority of galaxies reside in regions where $f\ll1$. A more quantitative assessment is shown in Fig. \ref{fig:pb_distribution}, where we plot in blue (left panel) the (differential) galaxy number counts distribution as a function of PB$_{\rm th}$, in green (central panel) the cumulative galaxy number counts distribution as a function of PB$_{\rm th}$ ($f>$PB$_{\rm th}$), and in orange (right panel) the reverse cumulative galaxy number counts distribution as a function of PB$_{\rm th}$ ($f<$PB$_{\rm th}$). This result clearly evidences that $\sim 85\%$ and $\sim 96\%$ of the galaxies lie in regions where $f<$PB$_{\rm th}=0.5$ and $f<$PB$_{\rm th}=0.8$, respectively. These findings allow us to realize that it is actually rather unlikely to obtain a statistically-significant SNR relying just on sources with high PB ($f\sim 1$), because the sample size is very small (few tens of sources) and the achieved SNR gain is not be sufficient to perform the detection of \HI{} signal, typically $<5$. 
Therefore, we limit our study at $f={\rm PB}_{\rm th}=0.6$, where the resulting sample is always constituted by more than $50$ galaxies. We anticipate that this does not represent a concern, as the PB correction is found in this work to have a negligible impact already at PB$_{\rm th}=0.6$ (see \S\ref{sec:pbcorr}), and is expected to has a lesser and lesser impact as PB$_{\rm th}\rightarrow 1$.  

\begin{figure*}
    \centering
    \includegraphics[width=18cm]{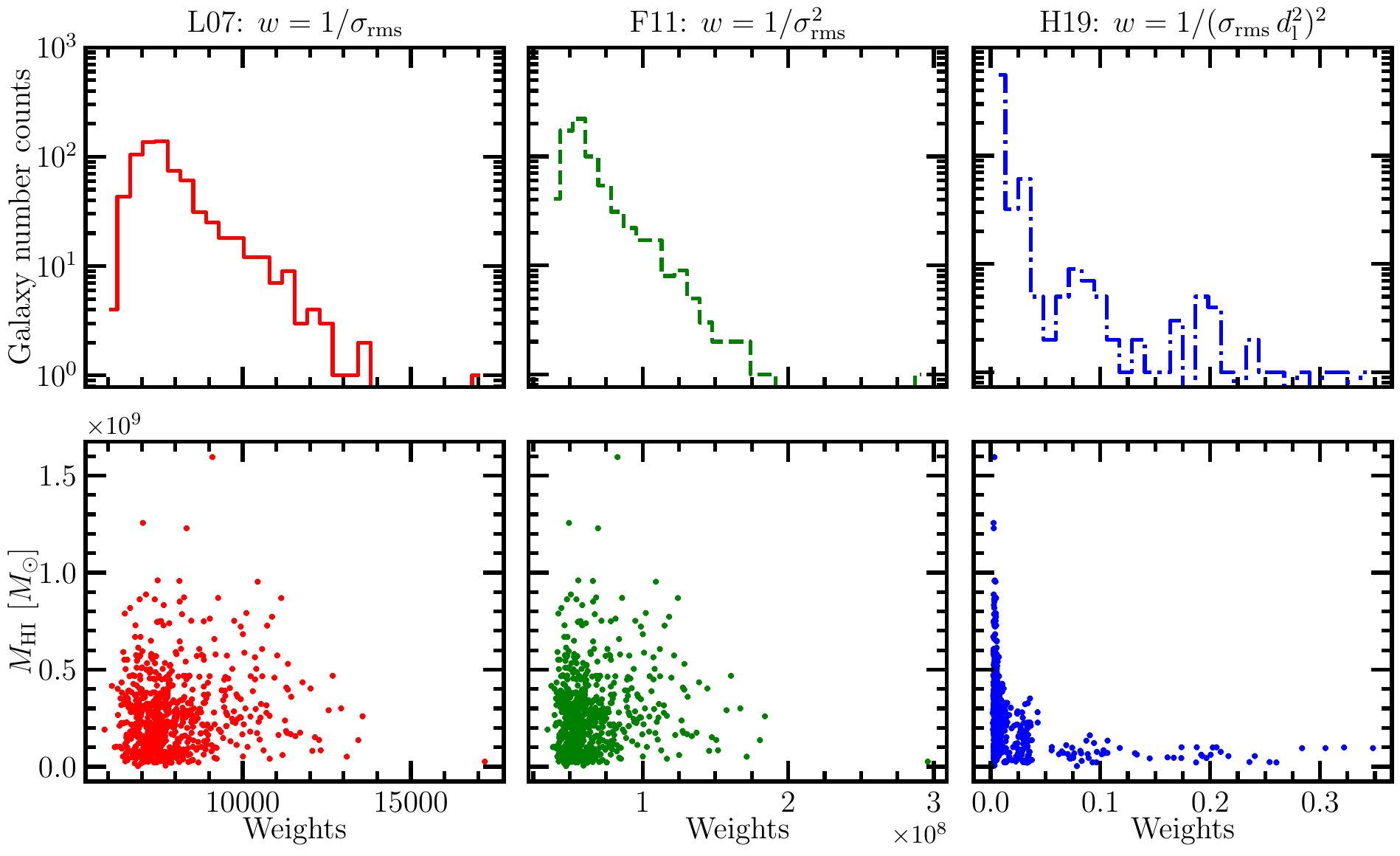}
    \caption{Top row: distribution of weights corresponding to mock galaxies, obtained following the L07 (left), F11 (centre), and H19 (right) weighting schemes. Bottom row: distribution of mock galaxies in the $M_{\rm HI}-$weights plane, with weights computed as in L07 (left), F11 (centre), and H19 (right).} 
    \label{fig:weights_distribution}
\end{figure*}

\subsubsection{$M_{\rm HI}$ deviations: unweighted case and PB correction} \label{sec:pbcorr}

The main results of this section are presented in Fig. \ref{fig:deltamass_wo_z_offset}. In panel (a), we show $\Delta M_{\rm HI}$ obtained by stacking in different versions of the cubes, as a function of PB$_{\rm th}$ and without any weighting schemes implemented (unweighted case, $w_i=1$). Here, gray shaded areas stand for $\pm 10\%$ (dark gray) and $\pm 20\%$ (light gray) errors, while red, green and blue shaded regions indicate error bars estimates through jackknife resampling, associated to lines of the same color. The red solid line shows the results for $\Delta M_{\rm HI}$ where $\braket{M_{\rm HI}}$ is obtained with stacking in the PB uncorrected noise-free cubes. Therefore, this line illustrates the impact of the primary beam correction alone, in the absence of noise. This allows us to study the theoretical intrinsic accuracy achieved by Eq. \ref{eq:pb_raw}. We observe that (i) the PB correction systematically tends to underestimate the true signal, and (ii) $\Delta M_{\rm HI}\sim -8\%$ at PB$_{\rm th}=0$ and then gradually increases (decreases in absolute values) until it reaches a nearly stationary state at mean deviation $\Delta M_{\rm HI}\sim -3\%$, although it is compatible with $\Delta M_{\rm HI}\sim 0$ within $1\sigma$. This result is in good agreement with the arguments presented in Appendix \ref{sec:appendix}. Therein, we show that the PB correction performed following Eq. (\ref{eq:pb_raw}) tends to induce a small overestimation of the $M_{\rm HI}$ signal -- of order just few per cent -- due to the finite sample size and intrinsic skewness of the $f$ and $M_{\rm HI}$ distributions, despite the fact that purely theoretical statistical argument tell that Eq. (\ref{eq:pb_raw}) provides unbiased results. The green dashed line presents the results for $\Delta M_{\rm HI}$ where $\braket{M_{\rm HI}}$ is obtained by stacking in flat noise-filled cubes, to estimate the effect of noise in absence of PB flux attenuation. In this case, the mean mass deviation obtained by averaging over PB$_{\rm th}$ bins is $\Delta M_{\rm HI}\sim 0$, as expected from the random Gaussian model for the noise that we adopted, with stochastic fluctuations as large as $\Delta M_{\rm HI}\sim 10\%$ in absolute value. This means that random noise does not introduce systematic uncertainties. Eventually, the blue dotted line shows the results for $\Delta M_{\rm HI}$ where $\braket{M_{\rm HI}}$ is obtained by stacking in PB uncorrected noise-free cubes, representing realistic data. We find that the HI signal is recovered with average mass deviation $\Delta M_{\rm HI}\sim -5\%$, with fluctuations within $\Delta M_{\rm HI}\sim -10\%$ and $\Delta M_{\rm HI}\sim 0$, except for the two last bins at PB$_{\rm th}>0.5$. In this case, the result can be understood as a superposition of the two previously discussed effects represented by the red solid and green dashed lines.

The final systematic mass deviation to be corrected a posteriori when performing stacking on real data, ideally at the correct redshift (i.e., very accurate and precise, with negligible uncertainties), is of order $\Delta M_{\rm HI}\sim -5\%$. However, we also notice that such a deviation is well compatible within $1\sigma$ uncertainties. This means that the systematic bias introduced by the PB correction is not statistically significant in the case studied here. 

\subsubsection{$M_{\rm HI}$: weighting schemes}

Panel (b) in Fig. \ref{fig:deltamass_wo_z_offset} shows $\Delta M_{\rm HI}$ as a function of PB$_{\rm th}$ for the different implemented weighting schemes, when both noise and PB flux attenuation are included in the mock datacubes. In practice, we extend the study we performed in panel (a) looking at the blue dotted line also to other weightings. The unweighted, L07 and F11 cases (red solid, green dashed, orange dashed-dotted lines, respectively) feature very similar results and are found to be substantially unbiased. Yet, we observe a slight underestimation of the signal of order $\Delta M_{\rm HI}\sim 5\%$, although within the uncertainties. Conversely, the distance-based H19 scheme causes a severe underestimation of the signal, of order $\Delta M_{\rm HI}\sim 40-50\%$. This has been already partially reported by \cite{Hu2019}, where the authors performed stacking on observations of $1895$ flux-limited optical galaxies acquired with the Westerbork Synthesis Radio Telescope (WSRT) and use the weighting $w_i=1/(\sigma^2_{\rm{rms}}\,d_l^\gamma)$, with $0<\gamma<4$. Their results (Fig. 5 and Table 1 in \citet{Hu2019}) highlight that the difference between the two extreme cases, $\gamma=0$ and $\gamma=4$, is a factor $\sim 2$ in $M_{\rm HI}$. The authors argue that the mismatch arises due to significant selection effects, where the case $\gamma=4$ gives too much weight to nearby galaxies, and $\gamma=0$ tends instead to be biased towards massive galaxies.

However, since their study is based on observations, the authors cannot determine which value of $\gamma$ maximises the accuracy of their measured average $M_{\rm HI}$ and conclude they use $\gamma=1$ as it maximizes SNR, and hence, minimizes the statistical error.

\begin{figure}
    \centering
    \includegraphics[width=\columnwidth]{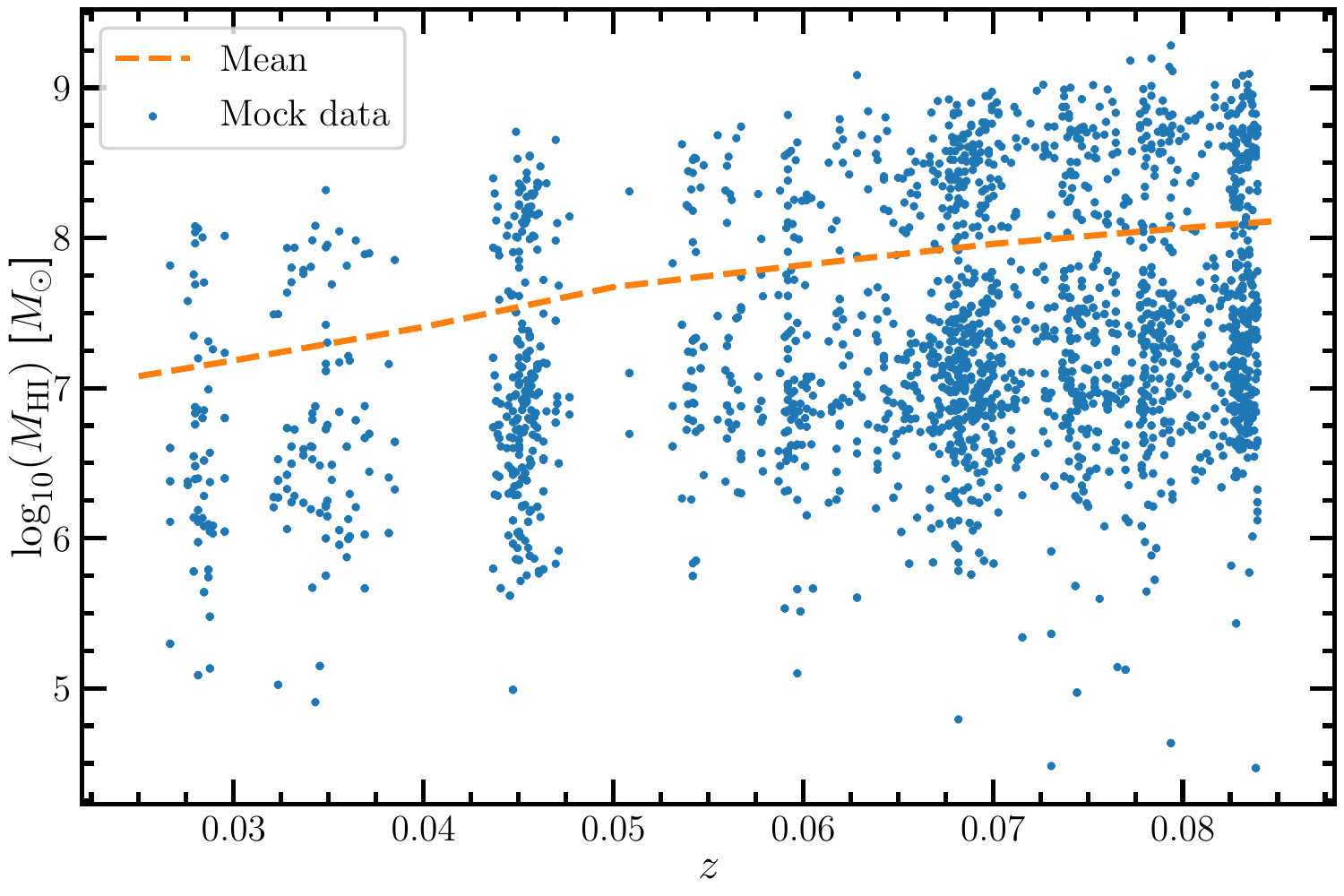}
    \caption{Distribution of $\log_{10}(M_{\rm HI})$ as a function of redshift. Blue points represent galaxies populating the mock data, while the orange dashed line stands for the mean trend.}
    \label{fig:mhi_vs_z}
\end{figure}

In this work, we are in position to investigate the origin of the aforementioned mismatch in mass between $\gamma=0$ and $\gamma=4$ cases. In fact, since we know the true $M_{\rm HI}$ associated to each galaxy in our sample, we can perform a detailed study of the problem. 


Fig. \ref{fig:weights_distribution} shows the weights distribution (top row) in the L07, F11 and H19 schemes (from left to right) and the two-dimensional distributions of galaxies in the $M_{\rm HI}$-weights plane in the three aforementioned cases. As expected from modelling the noise with a random sampling, there is no correlation between $M_{\rm HI}$ and weights in the L07 and F11 schemes, i.e. the two schemes in which weights are built as powers of noise rms. Conversely, $M_{\rm HI}$ and weights feature a clear correlation in the H19 case, where low-mass galaxies are systematically characterized by larger weights, confirming the argument by \cite{Hu2019}. This fact is what causes the underestimation of the mass signal we reported above. We can easily understand the origin of the correlation by looking at Fig. \ref{fig:mhi_vs_z}, showing the distribution of $M_{\rm HI}$ of mock galaxies in our sample as a function of redshift. Here, galaxies are identified by blue points and the mean $M_{\rm HI}(z)$ trend is represented as an orange dashed line. It turns out that galaxies are not uniformly distributed in $M_{\rm HI}$ across redshift. The reason behind this is primarily that our sample is heterogeneous since we exclude many bright detected galaxies and the detection limit is redshift-dependent. In fact, detected galaxies appear to have $\log_{10}(M_{\rm HI})\gtrsim 8$ at $z\sim 0.005$, while at $z=0.084$ detected galaxies have $\log_{10}(M_{\rm HI})\gtrsim 9$, i.e. there is $\sim 1$ dex difference in the $M_{\rm HI}$ detection limit at the two extremal redshifts. 
In realistic cases, even when there are no detected galaxies to be excluded from the stacking sample, a non-constant mean $M_{\rm HI}$ trend as a function of redshift can well be due to either intrinsic evolution of $M_{\rm HI}$ with redshift, or to selection effects if applied to a flux-limited sample. The latter can be e.g. due to the fact that more distant galaxies appear to be, on average, optically brighter and hence more \HI{}-massive (Malmquist bias). From a practical point of view, the origin of this effect is irrelevant: whenever $M_{\rm HI}$ is found to vary monotonically with redshift, a weighting based on distance will make systematically deviate the $\braket{M_{\rm HI}}$ estimate (with respect to the true $\braket{M_{\rm HI}}$) towards the average $\braket{M_{\rm HI}}$ of lower and lower $z$ galaxies the larger is $\gamma$. This becomes especially important when stacking is performed over a large $\Delta z$ interval. 

\begin{figure*}
    \centering
    \includegraphics[width=18cm]{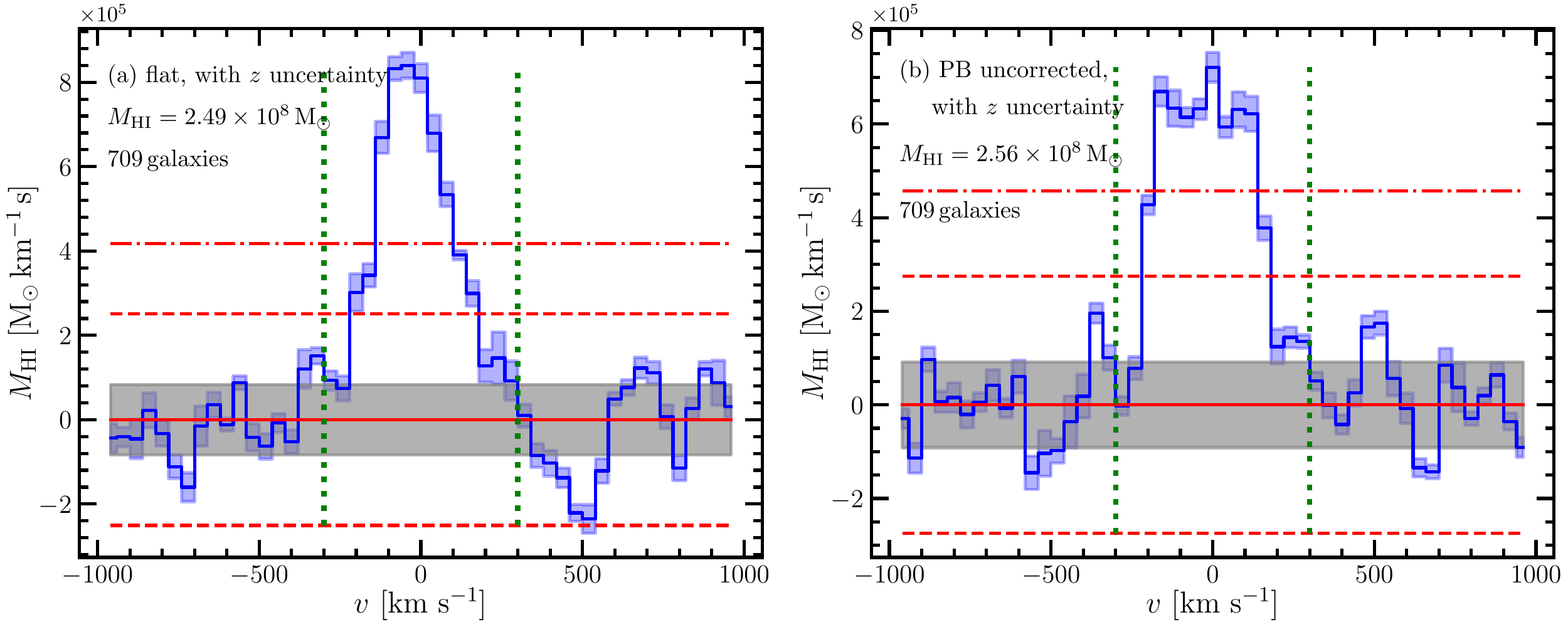}
    \caption{Stacked spectra obtained including all the available undetected sources in (a) flat cubes with redshift uncertainty, and (b) PB uncorrected cubes with redshift uncertainty. Blue lines indicate the resulting spectra, with corresponding shaded areas representing uncertainties obtained through jackknife resampling. Gray shaded areas, red dashed and red dashed-dotted lines stand for $1\sigma$, $3\sigma$ and $5\sigma$ noise rms levels.} 
    \label{fig:with_z_offset}
\end{figure*}

\begin{figure*}
    \centering
    \includegraphics[width=18cm]{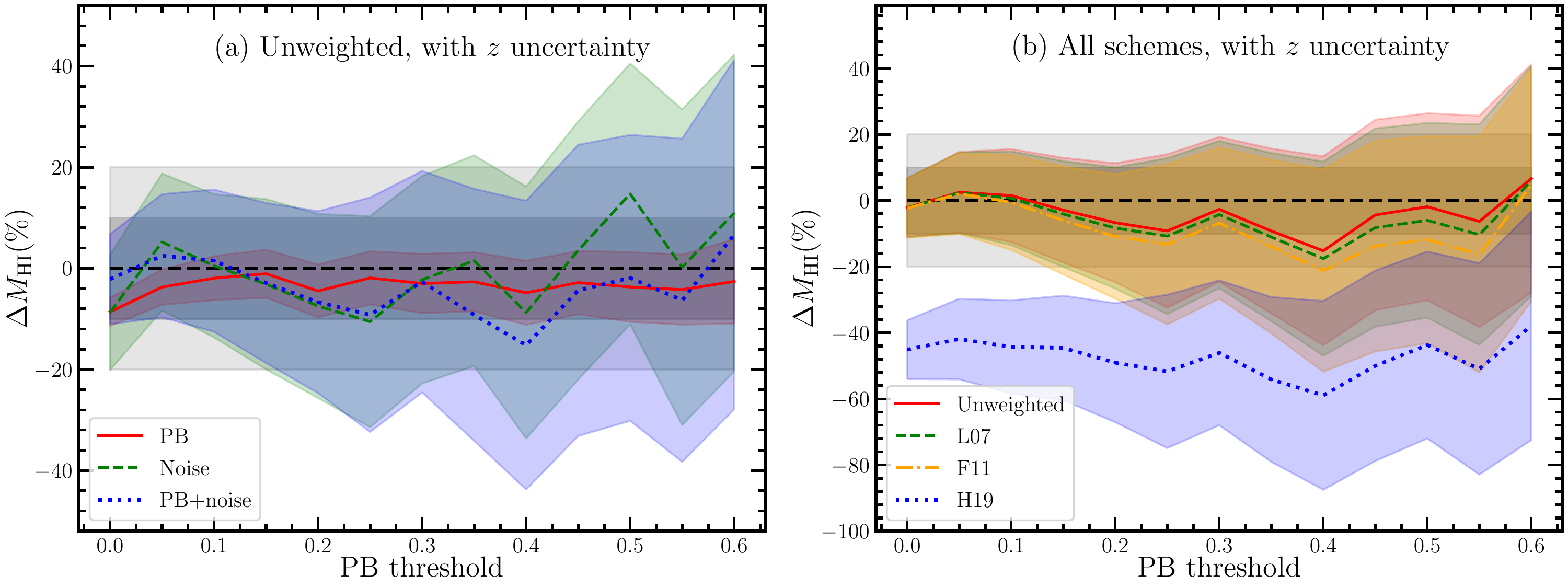}
    \caption{Percentage mass deviation $\Delta M_{\rm HI}$ as a function of PB$_{\rm th}$. Panel (a) shows the results of stacking in PB uncorrected noise-free cubes (red solid line), in flat noise-filled cubes (green dashed line) and in PB uncorrected cubes noise-filled (blue dotted line), representing the effects of PB correction alone, noise alone, and the combination of the two, respectively. No redshift offset and no weighting schemes (unweighted case) are applied. Panel (b) shows the results of stacking in PB uncorrected cubes noise-filled (applying PB correction in the presence of noise), with $4$ difference weightings: unweighted (red solid), L07 (green dashed), F11 (orange dashed-dotted), H19 (blue dotted). No redshift offset is applied. Shaded areas represent $1\sigma$ uncertainties around the curves they are associated to.} 
    \label{fig:deltamass_with_z_offset}
\end{figure*}

However, as mentioned, the true $\braket{M_{\rm HI}}$ may not be representative if the sample is flux-limited and a large redshift range is considered. We stress that, although the L07 and F11 weightings reproduce more accurately the mean $\braket{M_{\rm HI}}$ of our galaxy sample, the heterogeneity of the sample due to selection effects make the reference mean $\braket{M_{\rm HI}}$ potentially biased and not representative of the true average \HI{} content of the population. In this sense, distance-weighting provides a potential way to reduce the aforementioned bias, at the expense of having an additional degree-of-freedom. However, establishing which value of $\gamma$ yields physically representative results is not a trivial task, and depends on several factors involved in the definition of the sample. When relying on a volume-limited sample, instead than on a flux-limited one, distance-weighting will be much less characterized by $M_{\rm HI}$ variations with redshift in the galaxy sample due to selection effects, and one can use $\gamma$ as free parameter to find a compromise between the sensitivity and a potential $M_{\rm HI}$ variation with redshift. Yet, this does not guarantee that the resulting $\braket{M_{\rm HI}}$ is representative of the population, and depends on the intrinsic $M_{\rm HI}(z)$ evolution, which is unknown. We leave a careful investigation of this aspect for future works.

For what concerns this paper and other works analyzing flux-limited galaxy samples, we argue that special caution must be used in using distance-weighting schemes. 

\subsection{Stacking with redshift uncertainty} \label{sec:z_offsets}

In this section, we report results we obtain stacking at redshift $z'=z+\Delta z$. We randomly-sample redshift uncertainties $\Delta z$ from a Gaussian distribution with zero mean and standard deviation $\sigma_z=0.00037$, corresponding to a velocity interval $\Delta v\sim 105 $ km s$^{-1}$, to mimic the redshift uncertainty of the zCOSMOS survey \citep[][]{Lilly2007}, conducted with the VIMOS spectrograph at VLT.

Fig. \ref{fig:with_z_offset} shows the result of stacking applied to galaxies when $z$ offset described above is taken into account, in the flat (panel a) and PB uncorrected (panel b) cubes. In both cases, the effect of adding a redshift uncertainty to the galaxy sample consists in smearing the double horn profile and gaussianize the emission line \cite[see e.g.][]{Maddox2013,Jolly2020}. Furthermore, the resulting stacked emission is broadened, extending out to $\pm 500$ km s$^{-1}$, while in the case without $z$ offset the stacked mass profile drops sharply at $\pm 200$ km s$^{-1}$, as commented above.  

The main results of this section are reported in Fig. \ref{fig:deltamass_with_z_offset}, which features the same content as Fig. \ref{fig:deltamass_wo_z_offset}, though including $z$ uncertainty. We confirm the results found in the case without $z$ uncertainty. The unweighted, L07 and F11 schemes are consistent with an unbiased scenario, although with larger oscillations ($10-20\%$) around $\Delta M_{\rm HI}=0$. 
We also reiterate the result obtained with the H19 in the no-uncertainty analysis, i.e. distance-based weights are responsible for an average systematic deviation $\Delta M_{\rm HI}=-40-50\%$. Therefore, we find that spectroscopic redshift uncertainty does not introduce a bias on the estimate of $\braket{M_{\rm HI}}$, but rather slightly enlarges its uncertainty. 

\section{Summary and conclusions} \label{sec:conclusions}

In this work we have presented a novel framework to robustly assess the impact of common techniques adopted in 21-cm galaxy spectral stacking and estimate eventual corrections to be adopted a posteriori to recover the sought $\braket{M_{\rm HI}}$ signal. In particular, we have generated mock data mimicking $6$ interferometric pointings covering an area $\sim 6$ deg$^2$, equipped with realistic MeerKAT synthetized beam, primary beam, noise rms, spectral range and resolution. We have used these data products to study the impact of the commonly-adopted PB correction formula presented in Eq. (\ref{eq:pb_raw}) \citep{Gereb2013} and of weighting schemes widely-used in literature \citep{Lah2007,Fabello2011a,Delhaize2013,Hu2019}. We have also considered the impact of realistic spectroscopic redshift uncertainty, obtained by adding a random $z$ offset $\Delta z$ to each galaxy in our sample, obtained from a Gaussian distribution with zero mean and standard deviation $\sigma_z=0.00037$, to reproduce the features of the zCOSMOS survey \citep{Lilly2007}.


Our findings can be summarized as follows:
\begin{itemize}
    \item we find that the PB correction alone accounts for $\lesssim 8\%$ deviations on $\braket{M_{\rm HI}}$ measured through stacking, progressively decreasing when the sample is restricted by excluding galaxies located in regions where the normalized primary beam $f<\rm{PB}_{\rm th}$. We have limited our analysis to $\rm{PB}_{\rm th}=0.6$ as we have observed that the majority ($\sim 90\%$) of the galaxies in our sample are found at radial distance from the centre of the field-of-view such that $f<0.6$. At $\rm{PB}_{\rm th}=0.6$, the mass deviation is already almost negligible ($\sim 3\%$) and we argue that the PB correction should gradually have a lesser and lesser impact when $\rm{PB}_{\rm th}\rightarrow1$;
    \item we find that random noise (with Gaussian properties by construction) does not cause systematic deviations in the mass estimate, but rather accounts for random fluctuations around the mean signal of order $\Delta M_{\rm HI}\sim 5-10\%$;
    \item the unweighted, L07 and F11 weighting schemes, coupled to the PB correction, are found to be substantially unbiased, featuring just a slight underestimation of the signal of order $\Delta M_{\rm HI}\sim 5\%$, always compatible within $1\sigma$ with $\Delta M_{\rm HI}=0$. Yet, in realistic cases where the noise is not perfectly Gaussian, some larger deviations may come into play. However, at least at a theoretical level weighting spectra by their corresponding noise properties appears to be a robust procedure;
    \item distance-based weightings (H19, in this case) can account for significant deviations, $\Delta M_{\rm HI}\sim 40-50\%$ (in the case of our mock data), when the investigated galaxy population features a variation with redshift of the average underlying $M_{\rm HI}$. In the case of a flux-limited sample (as is the case of this paper), this can well be due to selection effects (e.g., Malmiquist bias). However, one should also notice that using the average mass as $\braket{M_{\rm HI}}$ estimate -- even though accurately recovered by the L07 and F11 schemes -- may not be representative of the global studied galaxy population and biased towards massive galaxies in a flux-limited sample. In a volume-limited sample, the signal underestimation due to the H19 weighting should be mitigated, at least partially, although such a weighting scheme may become degenerate with an intrinsic evolution of the amount of \HI{} in galaxies as a function of redshift. In particular, we notice that distance-weighting can hide an intrinsic $M_{\rm HI}(z)$ evolution also in a volume-limited sample of galaxies, giving systematically more weight to galaxies at lower $z$, especially if a large redshift interval $\Delta z$ is considered. In general, the freedom of choosing the exponent $\gamma$ may not yield a representative estimate for $\braket{M_{\rm HI}}$ either. We speculate that the possible variation on the estimates for $\braket{M_{\rm HI}}$ with different choices for $\gamma$ in a volume-limited sample can be conveniently exploited to mine the intrinsic evolution $M_{\rm HI}(z)$ in galaxies. We will explore this in future works. 
    
    We point out that stacking based on other properties, e.g. not $M_{\rm HI}$ but $M_{\rm HI}/L$ or $M_{\rm HI}/M_*$, being $L$ the galaxy luminosity in a given optical band and $M_*$ its stellar mass, can help alleviating the non-constant mass distribution with redshift due to selection effects;
    \item we propose, as a technique to be used as alternative to or in synergy with weights, to employ the symmetrized stacking to enhance the SNR by increasing the effective galaxy sample size, or, from another perspective, to increase the effective volume probed by observations. This novel procedure, introduced for the first time in this work (to the knowledge of the authors), is shown to be successful and unbiased. Symmetrized stacking is therefore particularly useful in cases where the stacked signal has low SNR, or even when it is undetected using traditional spectral or cubelet stacking;
\end{itemize}

We conclude that an appropriate framework to assess the accuracy of the stacking procedure, such as ours, is required to achieve adequate levels of accuracy on the way to the SKA. Even though the simulated datacubes presented in this work are designed to match the instrumental and technical features of MeerKAT observations, we argue that our procedure can be straightforwardly generalized to any radio telescope and array configuration. In particular, mock datacubes mimicking realistic observations conducted with a radio telescope, including models for the synthetized beam, for the primary beam and for the noise are recommended to assess the accuracy of the stacking setup and procedure.


\section*{Acknowledgements}

The authors warmly acknowledge the referee, Jean-Baptiste Jolly, for providing useful and insightful comments.
The authors also thank Tom Oosterloo, Natasha Maddox and Bradley Frank for useful discussions. F.S. acknowledges the support of the doctoral grant funded by the University of Padova and by the Italian Ministry of Education, University and Research (MIUR). G.R. acknowledges the support from grant PRIN MIUR 2017 - 20173ML3WW$\char`_$001. M.V. acknowledges financial support from the South African Department of Science and Innovation's National Research Foundation under the ISARP RADIOSKY2020 Joint Research Scheme (DSI-NRF Grant Number 113121) and the CSUR HIPPO Project (DSI-NRF Grant Number 121291). The authors also acknowledge the use of the ilifu cloud computing facility – \url{www.ilifu.ac.za}, a partnership between the University of Cape Town, the University of the Western Cape, the University of Stellenbosch, Sol Plaatje University, the Cape Peninsula University of Technology and the South African Radio Astronomy Observatory. The Ilifu facility is supported by contributions from the Inter-University Institute for Data Intensive Astronomy (IDIA – a partnership between the University of Cape Town, the University of Pretoria, the University of the Western Cape and the South African Radio Astronomy Observatory), the Computational Biology division at UCT and the Data Intensive Research Initiative of South Africa (DIRISA).

\section*{Data Availability}

The simulated datacubes underlying this article will be shared upon reasonable request to the corresponding author. 



\bibliographystyle{mnras}
\bibliography{lit} 



\appendix

\section{Intrinsic accuracy of primary beam correction}\label{sec:appendix}

In this section we present a more detailed analytical assessment of the accuracy of the primary beam correction formula presented in Eq. \ref{eq:pb_raw}:
\begin{equation} \label{eq:appendix1}
    S(\nu)=\frac{\sum_i f_i S_i'(\nu)}{\sum_i f_i^2} =\frac{\sum_i f_i^2 (S_i'(\nu)/f_i)}{\sum_i f_i^2}=\frac{\sum_i f_i^2 S_i(\nu)}{\sum_i f_i^2}  
\end{equation}
where $S(\nu)$, $S_i(\nu)$, $S'_i(\nu)$, and $f_i$ are the final stacked spectrum, the true and measured spectrum of the $i$th galaxy (i.e., the spectrum with flux reduced by $f_i$) and the primary beam power corresponding to the $i$th galaxy, respectively.

Assuming that both $S_i$ and $f_i$ are random variables, extracted from their corresponding probability distribution, Eq. \ref{eq:appendix1} reads:
\begin{equation}
 S(\nu)=\frac{\sum_i f_i^2 S_i(\nu)}{\sum_i f_i^2} = \frac{\left(\sum_i f_i^2 S_i(\nu)\right)/N_{\rm gal}}{\left(\sum_i f_i^2\right)/N_{\rm gal}} = \frac{\braket{f_i^2 S_i(\nu)}}{\braket{f_i^2}} \, .    
\end{equation}

Assuming that $S_i$ and $f_i$ are uncorrelated random variables, as there is no connection between galaxy fluxes and beam power, then:
\begin{equation}
  S(\nu) = \frac{\braket{f_i^2 S_i(\nu)}}{\braket{f_i^2}} = \frac{\braket{f_i^2} \braket{S_i(\nu)}}{\braket{f_i^2}} = \braket{S_i(\nu)} \, .
\end{equation}

Therefore, in principle one expects that the PB correction formula, on average, yields the correct underlying sought average flux.

However, this argument does not cover two important statistical aspects of the phenomenology of primary beam correction:
\begin{itemize}
    \item the variance (assuming Gaussian distribution, or more sophisticated indicators in the case of non-symmetric distributions) is in principle unknown;
    \item the sample size is finite and both the $S_i$ and $f_i$ distributions are highly skewed.
\end{itemize}

To thoroughly assess the impact of such points, we carry out numerical experiments on a synthetic distribution of galaxies. In particular, we randomly sample $S_i\curvearrowleft10^{\mathcal{N}(\mu=8.5,\sigma=1.0)}$ (in units $M_{\rm HI}$, see panel (c) of Fig. \ref{fig:galaxy_props}) and $f_i\curvearrowleft \left(\mathcal{U}[0,1]\right)^\alpha$, $\alpha>1$ (see left panel of Fig. \ref{fig:pb_distribution}), where $\curvearrowleft$, $\mathcal{N}$, and $\mathcal{U}$ stand for random sampling, Gaussian distribution and uniform distribution, respectively. For each set of $S_i$ and $f_i$ of size $N_{\rm gal}$, we  compute $S(\nu)$ and compare to the average $S_{\rm true}(\nu)$ obtained by construction, and compute percentage residuals as $R=100\times\left(S(\nu)/S_{\rm true}(\nu) - 1\right)$. We repeat the same experiment $n=1000$ times, and build a distribution of residuals. 

\begin{table}
    \centering
    \begin{tabular}{rcccc}
    \toprule
    $N_{\rm gal}$ & $\alpha$ & ${\rm med} \, (\%)$ & $\sigma \, (\%)$ & $\mathcal{P}(R<0) \, (\%)$\\ 
    \midrule
    $10^3$ & 1.5 & -2.7 & 25 & 55\\
    $10^3$ & 2 & -4.5 & 30 & 57\\
    $10^3$ & 3 & -6.7 & 37 & 58\\
    $10^3$ & 4 & -8.7 & 43 & 60\\
    \midrule
    $10^4$ & 1.5 & -1.1 & 12 & 52\\
    $10^4$ & 2 & -1.4 & 13 & 54\\
    $10^4$ & 3 & -2.5 & 17 & 57\\
    $10^4$ & 4 & -2-9 & 20 & 58\\
    \midrule
    $10^5$ & 1.5 & -0.2 & 4 & 52\\
    $10^5$ & 2 & -0.3 & 5 & 53\\
    $10^5$ & 3 & -0.6 & 7 & 55\\
    $10^5$ & 4 & -0.8 & 8 & 56\\
    \bottomrule
    \end{tabular}
    \caption{Result of the numerical experiments carried out to estimate the salient features (median, standard deviation and probability of getting negative residuals - third, fourth and fifth columns, respectively) of the distributions of $\Delta M_{\rm HI}$ residuals (see text), as a function of the sample size (first column) and exponent $\alpha$ controlling the skewness of distribution of the PB factors $f_i$. }
    \label{tab:residuals}
\end{table}

As diagnostic tools, we compute the median (med), the standard deviation ($\sigma$), and the probability of getting negative residuals $(\mathcal{P}(R<0))$, and tabulate them in Table \ref{tab:residuals}, as a function of the sample size $N_{\rm gal}$ and of the exponent $\alpha$. It turns out that systematically ${\rm med}<0$ and $\mathcal{P}(R<0)>50\%$. Moreover the larger $\alpha$ and the smaller $N_{\rm gal}$, the larger $|{
\rm med}|$, $\sigma$ and $\mathcal{P}(R<0)$. This means that the PB correction will increasingly underestimate the true signal with decreasing sample size and increasing skewness, due to the poor sampling of the tails of the distributions. In this scenario, the case represented by the simulated data we make use in the main body of the paper turns out to be a fairly representative case, consistent with $N_{\rm gal}\sim 700$ and $\alpha\sim 3$. Anyway, we notice in all the studied cases the $|\Delta M_{\rm HI}|$ due to PB correction is $\lesssim 10\%$, and hence represents a minor effect.



\bsp	
\label{lastpage}
\end{document}